\documentclass[12pt]{elsarticle}
\usepackage[latin9]{inputenc}
\usepackage{geometry}
\usepackage{amsmath}
\usepackage{amssymb}
\usepackage{graphicx}
\usepackage{esint}
\begin{document}

\title{The Inviscid, Compressible and Rotational, 2D Isotropic Burgers and
Pressureless Euler-Coriolis Fluids; Solvable models with illustrations}

\author[itp]{Ph. Choquard}

\ead{philippe.choquard@epfl.ch}

\author[lanl]{M. Vuffray}

\ead{vuffray@lanl.gov}

\address[itp]{Ecole Polytechnique Fédérale de Lausanne, ITP-SB-EPFL, Station 8,
CH-1015 Lausanne, Switzerland}

\address[lanl]{Center for Nonlinear Studies and Theoretical Division T-4, Los Alamos
National Laboratory, Los Alamos, NM 87545, USA}
\begin{abstract}
The coupling between dilatation and vorticity, two coexisting and
fundamental processes in fluid dynamics \citep[pp. 3, 6]{WMZ} is
investigated here, in the simplest cases of inviscid 2D isotropic
Burgers and pressureless Euler-Coriolis fluids respectively modeled
by single vortices confined in compressible, local, inertial and global,
rotating, environments. The field equations are established, inductively,
starting from the equations of the characteristics solved with an
initial Helmholtz decomposition of the velocity fields namely a vorticity
free and a divergence free part \citep[Sects. 2.3.2, 2.3.3]{WMZ}
and, deductively, by means of a canonical Hamiltonian Clebsch like
formalism \citep{Cleb.1}, \citep{Cleb.2}, implying two pairs of
conjugate variables. Two vector valued fields are constants of the
motion: the velocity field in the Burgers case and the momentum field
per unit mass in the Euler-Coriolis one. Taking advantage of this
property, a class of solutions for the mass densities of the fluids
is given by the Jacobian of their sum with respect to the actual coordinates.
Implementation of the isotropy hypothesis entails a radial dependence
of the velocity potentials and of the stream functions associated
to the compressible and to the rotational part of the fluids and results
in the cancellation of the dilatation-rotational cross terms in the
Jacobian. A simple expression is obtained for all the radially symmetric
Jacobians occurring in the theory. Representative examples of regular
and singular solutions are shown and the competition between dilatation
and vorticity is illustrated. Inspired by thermodynamical, mean field
theoretical analogies, a genuine variational formula is proposed which
yields unique measure solutions for the radially symmetric fluid densities
investigated. We stress that this variational formula, unlike the
Hopf-Lax formula, enables us to treat systems which are both compressible
and rotational. Moreover in the one-dimensional case, we show for
an interesting application that both variational formulas are equivalent. 
\end{abstract}
\maketitle

\section{Introduction\label{sec:Introduction}}

Consider the inviscid Burgers and pressureless Euler-Coriolis equations
in 2D, \eqref{eq:euler-coriolis} with $\omega=0$ and $\omega\neq0$
respectively, with initial velocity fields consisting of the gradients
of scalar potentials and of the orthogonal gradients of stream functions,
the said Helmholtz decomposition, and associated, here, to the models
of single vortices confined in compressible, local, inertial and global,
rotating environments. Our purpose is to study in these situations,
the coupling between dilatation and vorticity, two coexisting and
fundamental processes in fluid dynamics as emphasized by Wu et al.\ \citep[pp. 3,6]{WMZ}

The questions concerning the existence and uniqueness of explicit
solutions of these equations in general, are treated separately for
the rotational but incompressible case, in the monographs of Lions
\citep{PLL.1}, of Majda and Bertozzi \citep{MajBer}, of Marchioro
and Pulvirenti \citep{MarPul} and in the article of Shnirelman \citep{Shni},
and for the compressible but irrotational case, in the monograph of
Lions \citep{PLL.2} and in the article of Chen and Wang \citep{CheWan},
but, to our knowledge, not for the combined, Helmholtz like situation,
thus motivating the present paper. It is nevertheless worth quoting,
in particular, \citep[Sect. 2.2 and Ch. 8]{MajBer} and \citep[Sect. 10]{CheWan}
and the explicit solutions given by LI \citep{LI} for the inviscid
and incompressible Euler fluid in 2D and by Yuen \citep{Yuen} for
some anisotropic blowup solutions of the inviscid, compressible and
pressureless Euler equation in nD.

This paper is organized as follows. Section \ref{sec:Theoretical-Preambles}
presents theoretical preambles. It consists of four subsections. The
first one gives an inductive derivation of the relevant Burgers and
Euler equations in the sense of starting from the equations of their
characteristics with initial Helmholtz decomposition of their velocity
fields expressed in terms of Lagrangian variables, and moving up into
the Eulerian ones. Then, the formal, implicit solutions for the two
models are given. This is followed by a deductive derivation of the
field equations, a canonical Hamiltonian one, starting from Clebsch
Ansatz for the momentum field density of the fluid expressed in terms
of two pairs of canonically conjugated variables, one for the compressible
part and one for the rotational part of the fluid and from which the
equations of the characteristics are deduced. The next subject concerns
the fluid densities. Here, we exploit the fact that the Jacobian of
constant vector fields satisfies a continuity equation. Thus, starting
with our Helmholtz type constant velocity fields we choose our densities
to be proportional to the Jacobian of their sum, with a proportionality
constant having the dimension $\text{gc}\text{m}^{-2}\sec^{2}$, in
2D. It remains to implement our isotropy hypothesis, i.e. a radial
dependence of the potentials of the compressible and of the rotational
part of our fluids. The unexpected results are an additive contribution
of their compressible and rotational part with vanishing cross terms
and a simple algebraic expression for them namely: the derivative
of the square the vectorial velocity fields with respect to the square
of vectorial coordinates at time $t$, a result generalized to the
cases of $n$-dimensional, radially symmetric, at least twice differentiable
functions.

Section \ref{sec:Illustrations-and-variational-formulation} consists
of four subsections: two on solvable models, one on a variational
formula for these models and one on a one dimensional version of the
variational formula. The first illustration is that of a cylindrical
vortex in a local, compressible and inertial environment, i.e.\ a
Burgers case. The competition between dilatation and vorticity is
clearly demonstrated: according to the chosen initial conditions,
the density profile is regular for times larger than a critical one,
at which the density explodes, this critical time going to $-\infty$
for vanishing compressibility, as expected from the regularity properties
stated in the review articles cited above. In the second illustration,
an Euler-Coriolis case, we consider a cylindrical vortex in a global,
compressible and rotating environment. Here, the spiraling form of
the characteristics and their variation with the frequency of rotation
adds a second parameter leading to a new singularity, given analytically
but not illustrated numerically in this paper. However, for a given
frequency, the behavior of the density is similar to that of the first
illustration. It is the purpose of the third subsection to propose
a genuine variational formula inspired by a thermodynamical analogy
discovered between the present 2D isotropic situations and Weiss mean
field theory of Magnetism in 1D and which gives rise to unique measure
solutions for the densities investigated. As a fallout of the content
of this section, it is shown that the variational formula that we
propose, and based on a Maupertuis action instead of a Lagrangian
one, is equivalent to that of Hopf-Lax \citep[Sect. 3.3.2 and p.123]{Eva}
for the Burgers equation in 1D. This equivalence is exemplified at
hand of the one-dimensional Ising model in the mean field approximation.

Lastly Section \ref{sec:Further-developments} is devoted to the presentation
of further developments.

\section{Theoretical Preambles\label{sec:Theoretical-Preambles}}

\subsection{Equations of the characteristics and of the fluids\label{sub:Equations-of-the-characteristics}}

Let $\mathbf{r}=\left(r_{1},r_{2}\right)$, be the coordinates of
a test particle of mass $m$ in a 2 dimensional inertial reference
frame, $\mathbf{x=}\left(x_{1},x_{2}\right)$, its coordinates in
the rotating reference frame with frequency $\omega$ and let $\mathbf{O}(\omega t)=\begin{pmatrix}\cos\omega t & \sin\omega t\\
-\sin\omega t & \cos\omega t
\end{pmatrix}$ be the 2D orthogonal matrix. Let $\left\langle \mathbf{a}\mathbf{,}\mathbf{b}\right\rangle $
be the scalar product of the vectors $\mathbf{a}$ and $\mathbf{b}$,
and $\left\langle \mathbf{a,a}\right\rangle :=\mathbf{a}^{2}$, for
simplicity, and let $\mathbf{A}$ =$\begin{pmatrix}0 & -1\\
1 & 0
\end{pmatrix}$ be the anti-symmetric matrix such that the 2D vector product $\mathbf{a\wedge b=aAb}$.
With $\mathbf{x(}t)=\mathbf{O}(\omega t)\mathbf{r(}t)$, with $\dot{\mathbf{O}}(\omega t)=-\omega\mathbf{AO(}\omega t)$
and $\mathbf{\mathbf{\dot{x}}(t)=-}\omega\mathbf{AOr}+\mathbf{O\dot{r}}$\textbf{,
}we have

\begin{equation}
\dot{\mathbf{r}}=\mathbf{O}^{-1}(\mathbf{\mathbf{\dot{x}}+}\omega\mathbf{A\mathbf{x})},
\end{equation}
and the Lagrangian become
\begin{equation}
L(\dot{\mathbf{x}},\mathbf{x})=\frac{m}{2}\left\langle (\mathbf{\dot{x}}+\omega\mathbf{Ax),}(\mathbf{\dot{x}}+\omega\mathbf{Ax)}\right\rangle =\frac{m}{2}\left\langle \mathbf{\dot{r}},\mathbf{\dot{r}}\right\rangle =L^{\ast}(\mathbf{\dot{r}}).
\end{equation}
With the momenta $\mathbf{p=m(}\mathbf{\dot{x}}+\omega\mathbf{Ax),}$
and $\boldsymbol{\pi}=m\mathbf{\mathbf{\dot{r}}}$ the Hamiltonian
$H(\mathbf{p,x)}=\left\langle \mathbf{p,}\dot{\mathbf{x}}\right\rangle \mathbf{-}L(\dot{\mathbf{x}},\mathbf{x})=\frac{1}{2m}\mathbf{p}^{2}-\omega\mathbf{pAx}$
and $H^{\ast}(\boldsymbol{\pi})=\mathbf{\boldsymbol{\pi}}\mathbf{\dot{r}-}L^{\ast}(\mathbf{\dot{r}})$
become
\begin{equation}
H=\frac{1}{2m}\left\langle \mathbf{(p-}m\omega\mathbf{Ax),(p-m}\omega\mathbf{Ax)}\right\rangle -\frac{1}{2}m\omega^{2}\mathbf{\mathbf{x}}^{2}.
\end{equation}
and
\begin{equation}
H^{\ast}\mathbf{(\boldsymbol{\pi}\mathbf{)}}=\frac{1}{2m}\left\langle \mathbf{\boldsymbol{\pi},\boldsymbol{\pi}}\right\rangle .
\end{equation}
The resulting equations of motion are
\begin{equation}
\ddot{\mathbf{x}}+2\omega\mathbf{A}\mathbf{\dot{\mathbf{x}}}-\omega^{2}\mathbf{x=0},
\end{equation}
and
\begin{equation}
\mathbf{\ddot{\mathbf{r}}=0}.
\end{equation}
Inspection of the equation for $\mathbf{x(\mathrm{t})}$ shows that
its eigenvalues are degenerate thus explaining the spiraling nature
of the solutions. If $\mathbf{y}$ $=\left(y_{1},y_{2}\right)$ are
the initial coordinates, if $\phi\left(\mathbf{y}\right)$ is the
scalar potential associated to the compressible part of the fluid
and $\psi(\mathbf{y),}$ the stream function associated to its vorticity,
the initial velocity field $\mathbf{u}(\mathbf{y}),$ compatible with
Helmholtz decomposition \citep[Sects. 2.3.2, 2.3.3]{WMZ} then reads,
with $\mathbf{A}\nabla:=$ orthogonal gradient,
\begin{equation}
\mathbf{u}(\mathbf{y)=\nabla}\phi(\mathbf{y})\mathbf{+A\nabla}\psi(\mathbf{y}).
\end{equation}
The momentum per unit mass, $\boldsymbol{\pi}/m$, a constant vector
in the Coriolis case is denoted by
\begin{equation}
\mathbf{v}(\mathbf{y,}\omega):=\frac{\boldsymbol{\pi}}{m}\mathbf{=u(y)+}\omega\mathbf{Ay.}
\end{equation}
It follows that
\begin{equation}
\mathbf{r(y,}\omega,t)=\mathbf{y+}t\mathbf{v(y,}\omega\mathbf{)},
\end{equation}
and that
\begin{equation}
\mathbf{x}(\mathbf{y,}\omega,t\mathbf{)=O(}\omega t)(\mathbf{y+}t\mathbf{v(y,}\omega)),
\end{equation}
an explicit construction of the spirals.

It remains to pass from the Lagrangian to the Eulerian coordinates.
With $\mathbf{\dot{\mathbf{x}}=u:=u}(\mathbf{x,}\omega,t),$ $\mathbf{\ddot{\mathbf{x}}=\partial}_{t}\mathbf{u+}\left\langle \mathbf{u,\nabla}\right\rangle \mathbf{u,}$
we get, for the inviscid and pressureless Euler-Coriolis fluid,
\begin{equation}
\mathbf{\partial}_{t}\mathbf{u}+\left\langle \mathbf{u,\nabla}\right\rangle \mathbf{u+}2\omega\mathbf{Au-}\omega^{2}\mathbf{x=0.}\label{eq:euler-coriolis}
\end{equation}
For the Burgers cases, compressible and rotational, the terms containing
$\omega$ are omitted.

Let us point out here that, whereas the above derivation of the Euler-Coriolis
equation has followed an inductive path, in the sense that $\dot{\mathbf{x}}\rightarrow\mathbf{u(x},\omega,t)$,
and $\mathbf{\ddot{x}\rightarrow\partial}_{t}\mathbf{u+}\left\langle \mathbf{u,\nabla}\right\rangle \mathbf{u,}$
a deductive one is also feasible by means of a canonical Hamiltonian
Clebsch-like formalism \citep{Cleb.1}, \citep{Cleb.2} implying two
pairs of canonically conjugated field variables (3 in 3D), one pair
for the compressible part and one for the rotational part of the fluid,
a version presented in the next subsection.

It is appropriate to give, here, the formal, implicit solutions of
the equations corresponding to our two models. If $\mathbf{u}_{0}(\mathbf{y)}$
and $\mathbf{v}_{0}(\mathbf{y},\omega)$ designate, for simplicity
and for reasons of dimensionality, the two initial velocity fields,
(strictly speaking: initial velocity field and initial momentum field
per unit mass) and recalling that $\mathbf{x(}t\mathbf{)=O(}\omega t)\mathbf{r(t)}$,
then we have
\begin{equation}
\mathbf{u(x,t)=u}_{0}(\mathbf{y)=u}_{0}(\mathbf{x}-\mathbf{u(x,}t\mathbf{)}t),
\end{equation}
for the Burgers case ($\omega=0)$ and
\begin{equation}
\mathbf{v(r,}t,\omega)=\mathbf{v}_{0}(\mathbf{y,\omega)=v}_{0}(\mathbf{r-v(\mathbf{r,}}t\mathbf{,}\omega)t),
\end{equation}
for the Euler-Coriolis one $(\omega$$\neq$0).

Let us conclude this subsection in recalling that, with $\Delta$
being the Laplace operator, the dilatation field, $\Theta,$ is
\begin{equation}
\Theta\mathbf{=}\left\langle \mathbf{\nabla,v}\right\rangle =\Delta\phi,
\end{equation}
and the $z$ component of the vorticity field, $\mathbf{\Omega}$,
consisting of an intrinsic part and an extrinsic one, is
\begin{equation}
\mathbf{\Omega=\nabla}\wedge\mathbf{v=}\Delta\psi+2\omega.
\end{equation}

\subsection{Fluid-Mechanical Formulation\label{sub:Fluid-Mechanical-Formulation}}

In 2D, two canonically conjugate pairs of variables come into play:
$(\rho,\sigma)$ for the compressible part and $(\beta,\kappa)$ for
he rotational one. Clebsch's Ansatz for the total mass current, also
momentum density, is
\begin{equation}
\mathbf{j}=\rho\mathbf{\nabla}\sigma+\kappa\mathbf{\nabla}\beta.
\end{equation}
The pressureless Euler-Coriolis Hamiltonian is
\begin{align}
\hat{H}(\sigma,\beta;\rho,\kappa)= & \int_{R^{2}}d^{2}x\left(\frac{1}{2}\frac{\left\langle \mathbf{j,j}\right\rangle }{\rho}-\rho\omega\left\langle \mathbf{j,Ax}\right\rangle \right)\nonumber \\
= & \int_{R^{2}}d^{2}x\rho\left\langle \left(\mathbf{\nabla}\sigma+\frac{\kappa}{\rho}\mathbf{\nabla}\beta-\omega\mathbf{Ax}\right),\left(\mathbf{\nabla}\sigma+\frac{\kappa}{\rho}\mathbf{\nabla}\beta-\omega\mathbf{Ax}\right)\right\rangle \nonumber \\
 & -\frac{1}{2}\int_{R^{2}}d^{2}x\rho\omega^{2}\mathbf{x}^{2}.
\end{align}
The equations of motion are, in identifying $\mathbf{j/}\rho-\omega\mathbf{Ax=u,}$
the velocity field, and in setting $\frac{\kappa}{\rho}\mathbf{\nabla}\beta:=\mathbf{w}$,
\begin{equation}
\partial_{t}\sigma+\delta\hat{H}/\delta\rho=\partial_{t}\sigma+\frac{1}{2}\mathbf{u}^{2}-\left\langle \mathbf{u,w}\right\rangle -\frac{1}{2}\omega^{2}\mathbf{x}^{2}=0,
\end{equation}
\begin{equation}
\partial_{t}\rho-\delta\hat{H}/\delta\sigma=\partial_{t}\rho+\left\langle \mathbf{\nabla,}\rho\mathbf{u}\right\rangle =0,
\end{equation}
\begin{equation}
\partial_{t}\beta+\delta\hat{H}/\delta\kappa=\partial_{t}\beta+\left\langle \mathbf{u,\nabla}\beta\right\rangle =0,
\end{equation}
\begin{equation}
\partial_{t}\kappa-\delta\hat{H}/\delta\beta=\partial_{t}\kappa+\left\langle \mathbf{\nabla,}\kappa\mathbf{u}\right\rangle =0.
\end{equation}

Observe that $\beta$ is a constant of the motion and that $\kappa$,
in addition to $\rho$, satisfies the equation of continuity ;\ then,
setting $\kappa/\rho:=\alpha$ and $\mathbf{w:=}\alpha\mathbf{\nabla}\beta$
we notice that $\mathbf{w}$ can be identified with $\mathbf{A\nabla}\psi$
and we have that
\begin{equation}
\partial_{t}\alpha+\left\langle \mathbf{u,\nabla}\right\rangle \alpha=0.
\end{equation}

Thus, $\alpha$ and $\beta$ are two constants of the motion, also
called Clebsch parameters and the intersection of the surfaces $\alpha=c^{te}$and
$\beta=c^{te}$ gives vortex lines parallel to the z axes in our cylindrical
symmetry.In the illustrations, we have $\alpha$ $\mathbf{\nabla}\beta=\mathbf{A\nabla}\psi(y)=\mathbf{Ay}\psi'(y)/y$,
i.e $\beta(\mathbf{y)=\mathrm{\arctan\left(y_{2}/y_{1}\right)}=\varphi}$
and $\alpha=y\psi'(y)$ for $\psi(y)=\psi_{B}(y)$ and $\psi(y)=\psi_{EC}(y)$. 

We compute next the gradient of $\sigma$ with the aim of deriving
the pressureless Euler-Coriolis equation, an elementary operation
in the absence of intrinsic and extrinsic vorticity. We start from
\begin{equation}
\partial_{t}\mathbf{\nabla}\sigma+\left\langle \mathbf{u,\nabla}\right\rangle \mathbf{u+u\wedge\nabla\wedge u-\nabla}\left\langle \mathbf{u,w}\right\rangle -\omega^{2}\mathbf{x=0.}
\end{equation}
On the one hand, we have
\begin{equation}
\mathbf{\nabla}\left\langle \mathbf{u,w}\right\rangle =\left\langle \mathbf{u,\nabla}\right\rangle \mathbf{w+}\left\langle \mathbf{w,\nabla}\right\rangle \mathbf{u+u\wedge\nabla\wedge w+w\wedge\nabla\wedge u,}
\end{equation}
and, on the other hand, we have
\begin{align}
\partial_{t}\mathbf{w}:= & \partial_{t}(\alpha\mathbf{\nabla}\beta)\nonumber \\
= & \partial_{t}\alpha\mathbf{\nabla}\beta+\alpha\mathbf{\nabla}\partial_{t}\beta\nonumber \\
= & -\left\langle \mathbf{u,\nabla}\alpha\right\rangle \mathbf{\nabla}\beta-\alpha\mathbf{\nabla}\left\langle \mathbf{u,\nabla}\beta\right\rangle \nonumber \\
= & -\left\langle \mathbf{u,\nabla}\alpha\right\rangle \mathbf{\nabla}\beta-\alpha\left\langle \mathbf{\nabla u}\right\rangle \mathbf{\nabla}\beta-\alpha\left\langle \mathbf{\nabla}\beta,\mathbf{\nabla}\right\rangle \mathbf{u-}\alpha\mathbf{u\wedge\nabla\wedge\nabla}\beta\mathbf{-}\alpha\mathbf{\nabla}\beta\wedge\mathbf{\nabla\wedge u)}\nonumber \\
= & -\left\langle \mathbf{u,\nabla}\right\rangle \mathbf{w-}\left\langle \mathbf{w,\nabla}\right\rangle \mathbf{u-w\wedge\nabla\wedge u,}
\end{align}
having noticed that the term $\alpha\mathbf{u\wedge\nabla\wedge\nabla}\beta=\mathbf{0.}$
The important result is that 
\begin{equation}
\partial_{t}\mathbf{w=-\nabla}\left\langle \mathbf{u,w}\right\rangle +\mathbf{u\wedge\nabla\wedge w.}
\end{equation}
It follows that
\begin{equation}
\partial_{t}\left(\mathbf{\nabla}\sigma+\mathbf{w}\right)+\left\langle \mathbf{u,\nabla}\right\rangle \mathbf{u+u\wedge\nabla}\left(\mathbf{u-w}\right)-\omega^{2}\mathbf{x=0,}
\end{equation}
or, with $\mathbf{\nabla}\sigma+\mathbf{w=u+}\omega\mathbf{Ax}$ and
$\mathbf{u\wedge\nabla}\mathbf{\left(u-w\right)}=\mathbf{u\wedge\nabla}\left(\sigma-\omega\mathbf{Ax}\right)=-\omega\mathbf{u\wedge\nabla Ax}=2\omega\mathbf{Au}$,
the final result is
\begin{equation}
\partial_{t}(\mathbf{u+}\omega\mathbf{Ax)+}\left\langle \mathbf{u,\nabla}\right\rangle \mathbf{u+}2\omega\mathbf{Au-}\omega^{2}\mathbf{x=0.}
\end{equation}

In fact, it is a generalization of the equation established in the
first subsection since it applies to the cases where $\omega=\omega(t),$
a situation not considered in this paper. Other applications of Clebsch
canonical formalism to 2 and 3 D, compressible, self-interacting,
neutral and charged systems are given f.i. in \citep{ChoSs}, \citep[Sect. 2]{ChoMog},
and more, for magnetic and electromagnetic systems, leading, e.g.\ to
Euler-Lorentz and Euler-Maxwell equations are also possible.

So far we have been and still are able to use the elegant Clebsch
canonical formalism in our applications although the transformation
from Lagrangian to Eulerian variables is not canonical and our equations
of motion are not in a canonical form. Now, the fundamental fact that
Euler variables ($\rho$, $\mathbf{u}$ and $\mathbf{v}$ in our illustrations)
are not canonical variables has triggered an extraordinary rich development
of non canonical Hamiltonian theories, involving Lie-Poisson algebraic
structures and pioneered by Morrison and Greene in the eighties. Here,
it is most appropriate to quote Morrison's recent and very synthetic
presentation in \citep{Morr} where an exhaustive list of applications
and also a pertinent list of references are given.

\subsection{A class of solutions for the densities\label{sub:A-class-of-solutions-for-the-densities}}

Let us recall that, if $\mathbf{\xi(x,}t\mathbf{)}$ is a 2 dimensional
vector valued application such that each component is a constant of
the motion, i.e.\ $d\mathbf{\xi}/dt=\mathbf{\partial}_{t}\mathbf{\xi+}\left\langle \mathbf{u,\nabla}\right\rangle \mathbf{\xi=0}$,
then its Jacobian $J\mathbf{=\det(\partial}\xi_{i}/\partial x_{j})$
satisfies a continuity equation \citep[Section 3.3 and Apendix 5.2]{ChoMog}.
Furthermore, if $\mathbf{x}=\mathbf{x}(\mathbf{y},t)$ is the equation
of a characteristic, then we have that
\begin{equation}
J=\det(\partial\xi_{i}/\partial y_{j})\det(\partial x_{k}/\partial y_{l})^{-1}.
\end{equation}

In our cases, we have two constant fields: $\mathbf{u}$ and $\mathbf{v}$.
Considering first the Burgers case, and up to a proportionality constant,
of dimension $gcm^{-2}\sec^{2},$ set $=1,$ (the apparent dimension
of $\rho$ being $\sec^{-2}$), we define for the Burgers case,
\begin{equation}
\rho_{B}(\mathbf{x},t)=\mathbf{\det(\partial}u_{i}/\partial x_{j})=\det(\partial u_{i}/\partial y_{j})\det(\partial x_{k}/\partial y_{l})^{-1}.
\end{equation}
In the Euler-Coriolis case, and since $\det(\mathbf{O(}\omega t))=1,$
we define
\begin{equation}
\rho_{EC}(\mathbf{x,}\omega,t)=\rho_{EC}(\mathbf{r,}\omega,t)=\det(\partial v_{i}/\partial r_{j})=\det(\partial v_{i}/\partial y_{j})\det(\partial r_{k}/\partial y_{j})^{-1}.
\end{equation}
This class of densities might be identified as the Gelfand class \citep{Gelf}.
Whenever the method of characteristics is employed, more general solutions,
for any $\rho_{0}\left(\mathbf{y}\right)$ of positive type, would
be 
\begin{equation}
\rho(\mathbf{x,}t)=\rho_{0}\left(\mathbf{y}\right)\det(\partial x_{k}/\partial y_{l})^{-1}.
\end{equation}
The said Gelfand class will be used in what follows.It means that
the initial densities are proportional to $\det(\partial u_{i}/\partial y_{j})$
and to $\det(\partial v_{i}/\partial y_{j}).$

\subsection{Isotropic cases\label{sub:Isotropic-cases}}

In this subsection we consider radially-symmetric densities. Let $y=\left\vert \mathbf{y}\right\vert $,
$r=\left\vert \mathbf{r}\right\vert $, $x=\left\vert \mathbf{x}\right\vert $,
$u=\left\vert \mathbf{u}\right\vert ,$ and $v=\left\vert \mathbf{v}\right\vert .$
Thus, $\phi=\phi_{B}(y)$ or $\phi_{EC}(y,\omega)$, $\psi=\psi_{B}(y)$
or $\psi_{EC}(y,\omega)$ and for the Euler-Coriolis case we introduce
the effective potential $\Psi_{EC}(\omega,y)=\frac{1}{2}\omega y^{2}+\psi_{EC}(y,\omega).$
With $'$ and $''$ designating the first and second derivative with
respect to $y,$ we determine first the initial effective density.
It suffices to consider $\rho_{EC}(y,\omega).$ With $\partial_{y_{i}yj}^{2}f(y):=f_{ij}(\mathbf{y)}$
for any $f(y)$ twice differentiable, we have
\begin{align}
\rho_{EC}(y,\omega)= & v_{11}v_{22}-v_{12}v_{21}\nonumber \\
= & (\phi_{_{EC,}11}-\Psi_{EC,21})(\phi_{EC,22}+\Psi_{EC,12})-\left(\phi_{EC,12}-\Psi_{EC,22}\right)\left(\phi_{EC,21}+\Psi_{EC,11}\right).
\end{align}
A detailed calculation shows that, in the isotropic case, the cross
terms in $\phi_{EC,ij}\Psi_{EC,kl}$ cancel out. Thus we get
\begin{equation}
\rho_{EC}(y,\omega)=\det(\phi_{EC,ij})+\det\left(\Psi_{EC,ij}\right).
\end{equation}
It remains to calculate one of these determinants, for example\ and
generically,
\begin{align}
\det(\phi_{ij}\left(y\right))= & \left(\frac{y_{2}^{2}}{y^{3}}\phi'\left(y\right)+\frac{y_{1}^{2}}{y^{2}}\phi''\left(y\right)\right)\left(\frac{y_{1}^{2}}{y^{3}}\phi'\left(y\right)+\frac{y_{2}^{2}}{y^{2}}\phi''\left(y\right)\right)\nonumber \\
 & -\left(\frac{y_{1}y_{2}}{y^{2}}\right)\left(-\frac{\phi'\left(y\right)}{y}+\phi''\left(y\right)\right)^{2}\nonumber \\
= & \left(\frac{y_{1}^{4}+2y_{1}^{2}y_{2}^{2}+y_{2}^{4}}{y^{5}}\right)\phi'(y)\phi''(y)\nonumber \\
= & \frac{1}{y}\phi'(y)\phi''(y)=\frac{d\left(\phi'(y)\right)^{2}}{d\left(y^{2}\right)}.
\end{align}
It follows that
\begin{equation}
\rho_{B}(y)=\frac{d\left(\phi'_{B}(y)\right)^{2}}{d\left(y^{2}\right)}+\frac{d\left(\psi'_{B}(y)\right)^{2}}{d\left(y^{2}\right)}=\frac{d\left(\mathbf{u}^{2}(y)\right)}{d\left(y^{2}\right)}.
\end{equation}
and that
\begin{equation}
\rho_{EC}(y,\omega)=\frac{d\left(\phi_{EC}'(y,\omega)\right)^{2}}{d\left(y^{2}\right)}+\frac{d\left(\Psi_{EC}'(y,\omega)\right)^{2}}{d\left(y^{2}\right)}=\frac{d\left(\mathbf{v}^{2}(y,\omega)\right)}{d\left(y^{2}\right)}.
\end{equation}

In summary, the isotropy hypothesis and the symmetries of the velocity
fields result in the fact that the Jacobian of a sum equals the sum
of the Jacobians and in a simple formula for them. In fact this formula
can be generalized to the cases of $n$-dimensional radially-symmetric,
at least twice-differentiable functions $f\left(y\right)$. Indeed,
let $e_{i}=\frac{y_{i}}{y}$, $i=1,\dots,n$, be the director cosines
of the vector $\mathbf{y}$ and let the matrix elements $f_{ij}=\frac{f'\left(y\right)}{y}\delta_{ij}+\left(\frac{-f'\left(y\right)}{y}+f''\left(y\right)\right)e_{i}e_{j}$,
then the Jacobian is $J=\det\left(f_{ij}\left(y\right)\right)=\left(\frac{f'\left(y\right)}{y}\right)^{n-1}f''\left(y\right)=\frac{d}{d\left(y^{n}\right)}\left(f'\left(y\right)\right)^{n}$,
owing to the vanishing of all sub-determinants with elements $e_{i}e_{j}$.

We proceed with the evaluation of the Jacobian 
\begin{equation}
\det(\partial x_{k}/\partial y_{l})=1+t\Delta\phi(y)+\frac{t^{2}}{y}(\phi'(y)\phi''(y)+\psi'(y)\psi''(y)),
\end{equation}
 and 
\begin{equation}
\det\left(\partial r_{k}/\partial y_{l}\right)=1+t\Delta\phi\left(y\right)+\frac{t^{2}}{y}\left(\phi'(y)\phi''(y)+\omega^{2}+\omega\Delta\psi(y)+\psi'(y)\psi''(y)\right).
\end{equation}
 We find
\begin{equation}
\det(\partial x_{k}(y,t)/\partial y_{l})=\frac{\partial\left(\mathbf{x}^{2}(y,t)\right)}{\partial\left(y^{2}\right)},
\end{equation}
and
\begin{equation}
\det(\partial r_{k}(\mathbf{y,}\omega,t)/\partial y_{l})=\frac{\partial\left(\mathbf{r}^{2}(y,\omega,t)\right)}{\partial\left(y^{2}\right)}.
\end{equation}
It follows that
\begin{equation}
\rho_{B}(x,t)=\frac{d\left(\mathbf{u}^{2}(y)\right)}{d\left(y^{2}\right)}\left(\frac{\partial\left(\mathbf{x}^{2}(y,t)\right)}{\partial\left(y^{2}\right)}\right)^{-1}=\frac{\partial\left(\mathbf{u}^{2}(x,t)\right)}{\partial\left(\mathbf{x}^{2}\right)},
\end{equation}
and, similarly, that
\begin{equation}
\rho_{EC}(r,\omega,t)=\frac{d\left(\mathbf{v}^{2}(y,\omega)\right)}{d\left(y^{2}\right)}\left(\frac{\partial\left(\mathbf{r}^{2}(y,\omega,t)\right)}{\partial\left(y^{2}\right)}\right)^{-1}=\frac{\partial\left(\mathbf{v}^{2}\mathbf{(}r,\omega,t)\right)}{\partial\left(\mathbf{r}^{2}\right)}.
\end{equation}

\section{Illustrations and variational formulation\label{sec:Illustrations-and-variational-formulation}}

Two illustrations are proposed in this section and their results interpreted
in terms of a variational formula inspired by thermodynamical analogies.

\subsection{A cylindrical vortex in a compressible, finite and inertial, environment:
a Burgers case\label{sub:A-cylindrical-vortex}}

The initial velocity fields are split into a compressible and a rotational
part. The corresponding mass densities are such that the total mass
is prescribed. With $-\pi/2\leq\tau\leq\pi/2$, $0<\lambda\leq\mu,$
for the range parameters, we choose, in omitting in the functions
an explicit quotation of the parameters, unless otherwise convenient
and re-calling that $\mathbf{r=x}$, since $\omega=0,$
\begin{equation}
\phi_{B}'(y)=\frac{\lambda y}{\sqrt{(1+\left(\lambda y\right)^{2})}}\cos\tau,
\end{equation}
and, allowing negative amplitude for the rotational part,
\begin{equation}
\psi_{B}'\left(y\right)=\frac{\mu y}{\sqrt{(1+\left(\mu y\right)^{2})}}\sin\tau.
\end{equation}
Designating the corresponding initial field densities by $\rho_{\phi_{B}}(y)$
and$\rho_{\psi_{B}}(y)$ and their sums by $\rho_{B}(y),$ we have
\begin{equation}
\rho_{B}(y):=\rho_{\phi_{B}}+\rho_{\psi_{B}}=\frac{\left(\lambda\cos\tau\right)^{2}}{\left(1+\left(\lambda y\right)^{2}\right)^{2}}+\frac{\left(\mu\sin\tau\right)^{2}}{\left(1+\left(\mu y\right)^{2}\right)^{2}}.
\end{equation}

\begin{figure}

\begin{centering}
\includegraphics[width=10cm]{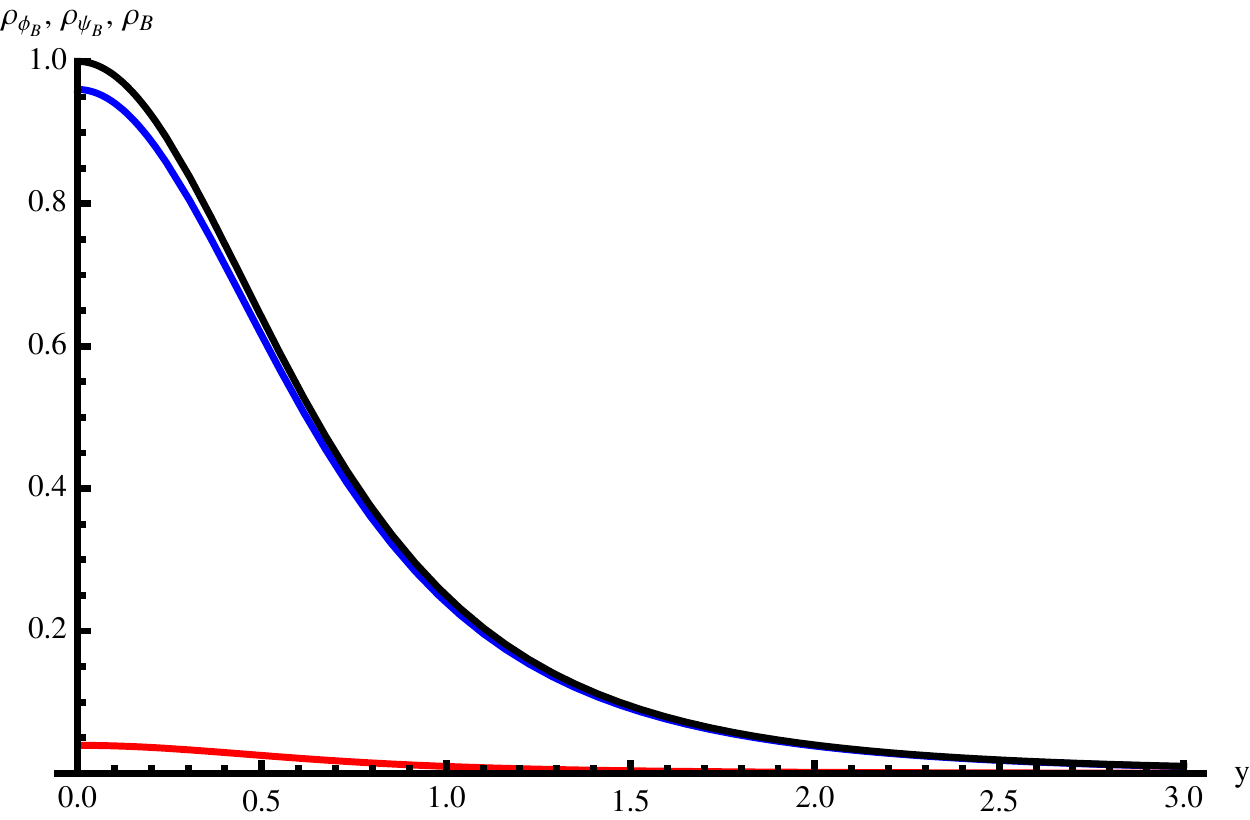}\protect\caption{\label{fig:bgraph1}Initial density of the Burgers case. The curves
represent $\rho_{\phi_{B}}$ (blue), $\rho_{\psi_{B}}$ (red) and
$\rho_{B}$ (black).}

\par\end{centering}

\end{figure}
They are displayed in Figure \ref{fig:bgraph1} for $\lambda=\mu=1$
and $\tau=0.2,$ a particular case considered later.

This model represents a cylindrical vortex with positive or negative
amplitude, in a compressible, finite and inertial environment. Clearly,
the total mass is, per unit height,
\begin{equation}
\pi{\displaystyle \int\limits _{0}^{\infty}}dy^{2}\rho_{B}(y,0)=\pi\left(\phi_{B}'\left(\infty\right)^{2}+\psi_{B}'\left(\infty\right)^{2}\right)=\pi.
\end{equation}
We have also
\begin{equation}
\mathbf{u}(y)^{2}=\phi_{B}'(y)^{2}+\psi_{B}'\left(y\right)^{2},
\end{equation}
and
\begin{equation}
\mathbf{x}(y,t)^{2}=\left(y+t\phi_{B}'(y)\right)^{2}+\left(t\psi_{B}'\left(y\right)\right)^{2}.
\end{equation}

Consider the domain of regularity for the densities. Since $\rho_{B}(y)>0$
for $y<\infty$, the condition is, in omitting again the explicit
dependence upon the parameters $\lambda$, $\mu$, $\tau$,
\begin{equation}
\frac{\partial\left(\mathbf{x}^{2}(y,t)\right)}{\partial\left(y^{2}\right)}=(1+t\phi_{B}'(y)/y)\left(1+t\phi_{B}''(y)\right)+t^{2}\psi_{B}'(y)\psi_{B}''(y)/y>0.
\end{equation}
The two roots in the time variable of the above relation set equal
to zero are, with $\phi_{B}'(y)/y+\phi_{B}''(y)=\Delta\phi_{B}(y),$
\begin{equation}
t_{\pm}(y)=\left(-\Delta\phi_{B}(y)\pm\left(\left(\Delta\phi_{B}(y)\right)^{2}-4\rho_{B}(y)\right)^{1/2}\right)/2\rho_{B}(y),
\end{equation}
and the positivity condition implies that the discriminant $D_{B}(y)=\left(\Delta\phi_{B}(y)\right)^{2}-4\rho_{B}(y,0)<0.$

As illustration, let $\mu=\lambda=1$, i.e. a one-dimensional subspace
$(\tau)$ of the three dimensional parameter space $(\lambda,\mu,\tau)$.
In this case, $\mathbf{u}^{2}=\mathbf{y}^{2}\left(1+y^{2}\right)^{-1}$
or $\mathbf{y}^{2}=\mathbf{u}^{2}\left(1-\mathbf{u}^{2}\right)^{-1}$,
$0\leq u\leq1$ and 
\begin{equation}
\mathbf{x}^{2}(\mathbf{u}^{2},t,\tau)=\mathbf{u}^{2}\left(1-\mathbf{u}^{2}\right)^{-1}+2t(\cos\tau)\mathbf{u}^{2}\left(1-\mathbf{u}^{2}\right)^{-1/2}+t^{2}\mathbf{u}^{2}.
\end{equation}
It follows that in investigating the inverse density $\rho_{B}(\mathbf{u}^{2},t,\tau)^{-1}$
$=\frac{\partial\left(\mathbf{x}^{2}\right)}{\partial\left(\mathbf{u}^{2}\right)}$
and requiring that $\rho_{B}(\mathbf{u}^{2},t,\tau)^{-1}>0,$ we can
determine the domain of regularity of the solutions, i.e.
\begin{equation}
\rho_{B}(\mathbf{u}^{2},t,\tau)^{-1}=(1-\mathbf{u}^{2})^{-2}+t(\cos\tau)(2-\mathbf{u}^{2})(1-\mathbf{u}^{2})^{-3/2}+t^{2}>0.
\end{equation}

There is a one parameter family of critical points $t_{c}(\tau)<0$
at which the two roots of this quadratic equation in $t$ (set $=0$)
coincide. Setting the discriminant of this equation $=0$ gives an
equation for the critical values of the velocity, namely
\begin{equation}
\left(\cos\tau\right)^{2}=4\frac{1-\mathbf{u}_{c}^{2}}{\left(2-\mathbf{u}_{c}^{2}\right)^{2}}.
\end{equation}
Introducing the angular variable $\theta$ such that $1-\mathbf{u}_{c}^{2}=\left(\tan\theta\right)^{2}$
results in $\left(\cos\tau\right)^{2}=\left(\sin\left(2\theta\right)\right)^{2}$.
The relevant solution is $2\theta=\pi/2-\left\vert \tau\right\vert .$
It follows that $0\leq\theta\leq\pi/4$ and thus, $\mathbf{u}_{c}^{2}(\tau)=1-\left(\tan\left(\pi/4-\left\vert \tau/2\right\vert \right)\right)^{2}.$

For the critical times we have, in setting the discriminant $=0$,
$t_{c}=-\frac{1}{1-\mathbf{u}_{c}^{2}}$, or
\begin{equation}
t_{c}(\tau)=-1/\left(\tan\left(\pi/4-\left\vert \tau\right\vert /2\right)\right)^{2}.
\end{equation}
It is interesting to notice that $t_{c}(0)=-1$ and $u_{c}(0)=0$,
for the purely compressible case, whereas $t_{c}(\pm$ $\pi/2)=-\infty$
and $u_{c}(\pi/2)=1$ for the purely rotational one: this means that,
in the coexisting processes of dilatation and vorticity, the domain
of regularity increases with increasing vorticity \citep{WMZ}.
\begin{figure}

\begin{centering}
\includegraphics[width=10cm]{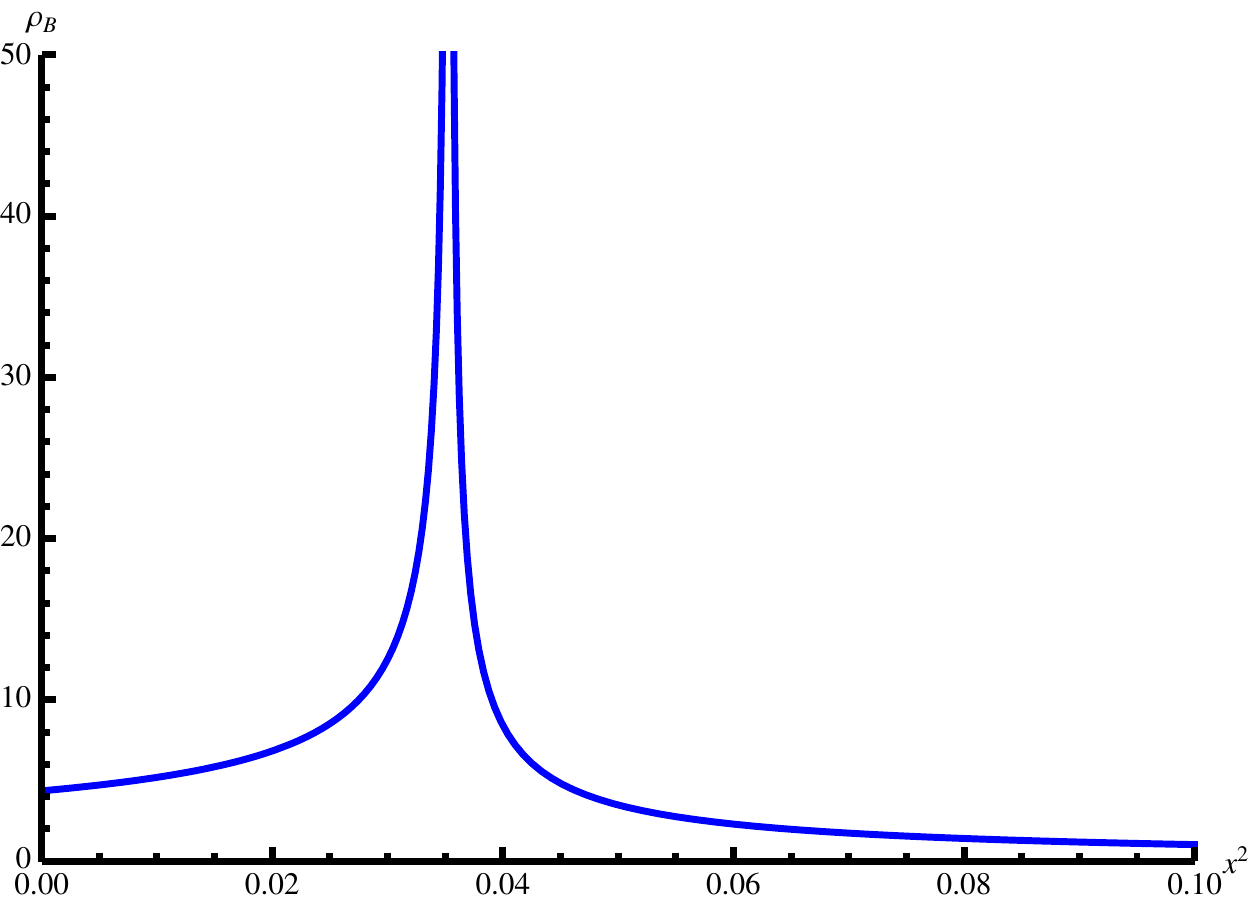}\protect\caption{\label{fig:bgraph6}Critical density of the Burgers case.}

\par\end{centering}

\end{figure}

Let us take the example considered above. For $\lambda=\mu=1$ and
$\tau=0.2,$ one finds $t_{c}\thickapprox-1.4$$,$ $y_{c}\thickapprox0.7$,
$x_{c}\thickapprox0.2$ and $u_{c}^{2}\thickapprox0.4.$ The density
is regular for $t\geq t_{c}$, with its maximum at the origin until
$t\approx-0.83$, when its curvature changes sign, from concave to
convex, signaling the onset of a maximum emerging from the origin
and culminating to its blow up at $t=t_{c}.$ Figure \ref{fig:bgraph6}
shows $\rho_{B}(\mathbf{x}^{2},t_{c}).$ Algebraic singularities in
$\mathbf{x}^{2}$ and in $\mathbf{u}^{2}$ are found to be an asymmetric
one for $\rho_{B}(\mathbf{x}^{2},t_{c})^{-1}\thicksim(\alpha_{+}\chi(\mathbf{x}^{2}-\mathbf{x}_{c}^{2})+\alpha_{-}\chi(\mathbf{x}_{c}^{2}-\mathbf{x}^{2}))$,
$\chi(\xi)$ being the characteristic function, $\alpha_{+}$, $\alpha_{-}$,
positive amplitudes, and a quadratic one $\thicksim(\mathbf{u}^{2}-\mathbf{u}_{c}^{2})$$^{2}$
for $\rho_{B}(\mathbf{u}^{2},t_{c})^{-1}.$ 
\begin{figure}

\begin{centering}
\includegraphics[width=8cm]{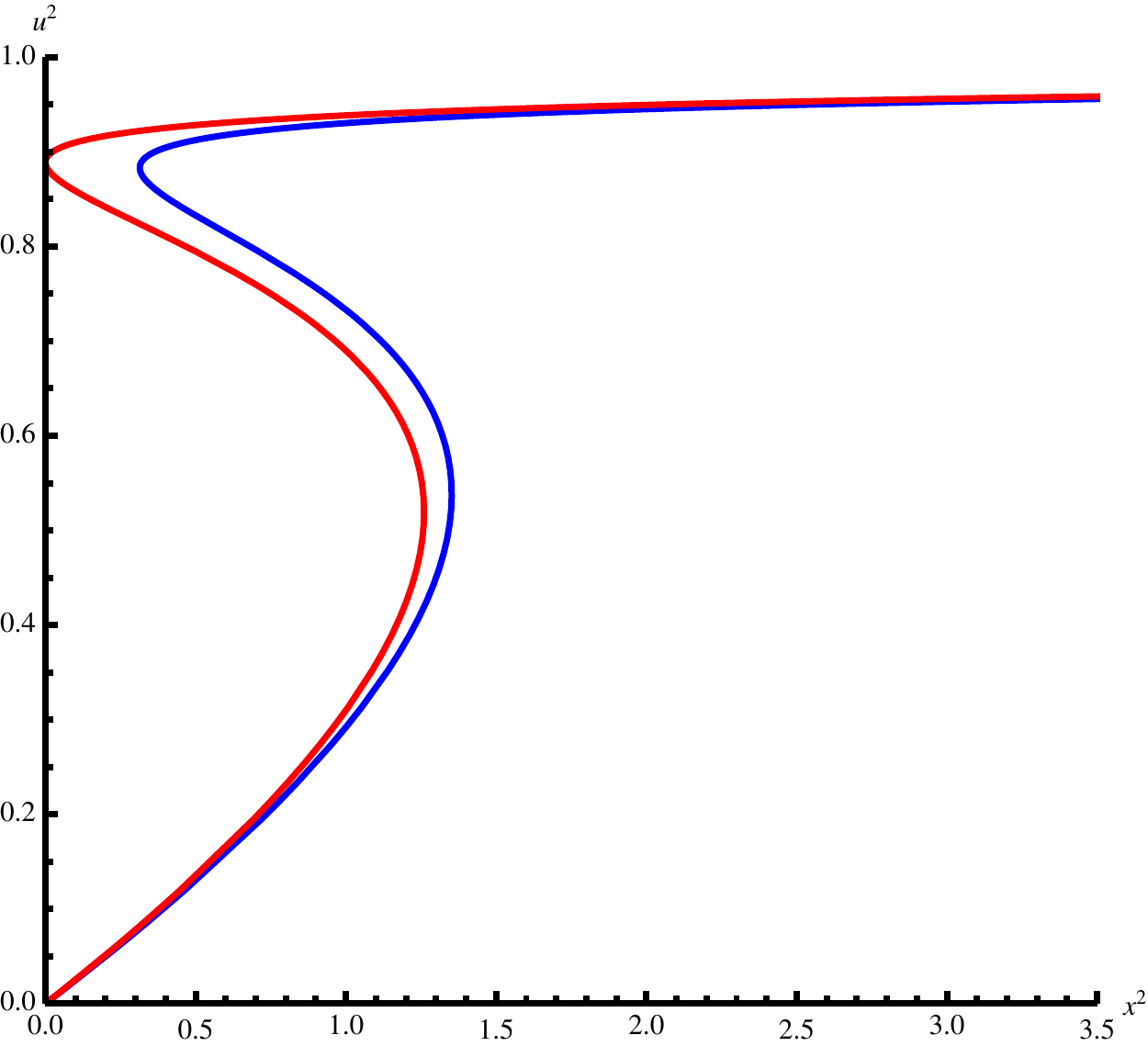}\protect\caption{\label{fig:bgraph4}Weiss analogy of the Burgers case for subcritical
densities at $\tau=0$ (red) and $\tau=0.2$ (blue).}

\par\end{centering}

\end{figure}

The analysis of the behavior of this model for times $<t_{c}(\tau)$
will be postponed after the presentation of the second illustration.
As a hint, Figure \ref{fig:bgraph4} shows the graphs of $\mathbf{u}^{2}(\mathbf{x}^{2},t)$
for $t=t_{<}=-3$, and, as before, $\lambda=\mu=1,$ $\tau=0.2$ and
also $\tau=0$, corresponding to the purely compressible case. 
\begin{figure}
\begin{centering}
\includegraphics[width=8cm]{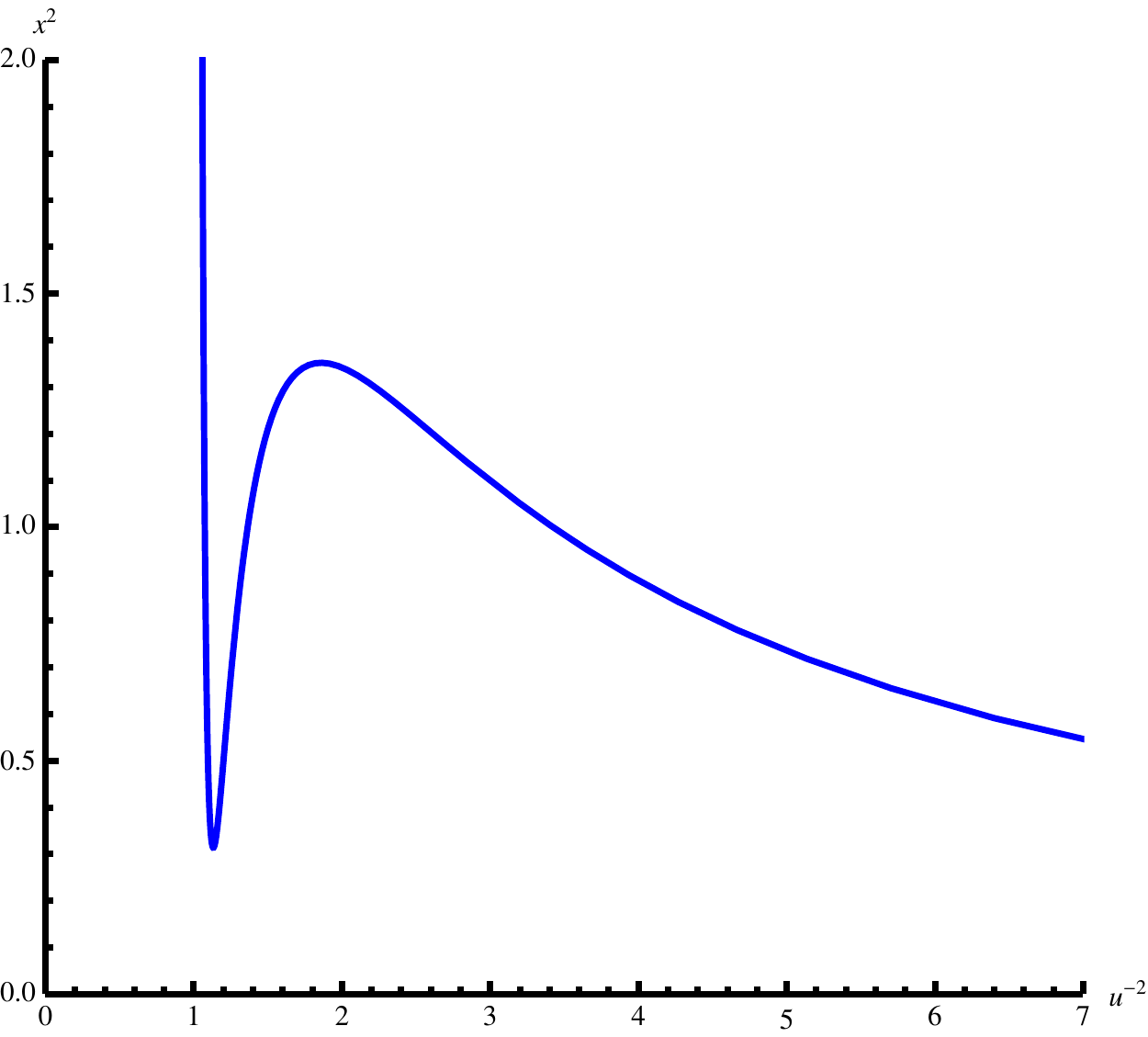}\protect\caption{\label{fig:bgraph5}Van der Waals analogy of the Burgers case.}

\par\end{centering}

\end{figure}
The analogy $\mathbf{u}^{2}\backsim\left(\text{magnetization}\right)^{2}$,
$\mathbf{x}^{2}\backsim\left(\text{{\rm magnetic field}}\right)^{2}$,
$-t\backsim\left(\text{coupling const./temperature}\right)$, with
Weiss's theory of magnetism is particularly striking if, for the $\tau=0$
case, we fold in the first quadrant the curve: magnetization versus
magnetic field \citep[Eqn. 23]{ChoMF}. Alternatively, the graph of
$\mathbf{x}^{2}(\mathbf{u}^{-2},t_{<})$, shown in Figure \ref{fig:bgraph5},
with $\mathbf{x}^{2}$$\backsim$pressure, $\mathbf{u}^{-2}$$\backsim$volume,
evokes a van der Waals loop, with $\mathbf{u}^{-2}\geq1$ mimicking
a hard core repulsion. However, no coexistence of phases being expected
here, the first analogy is favored.

\subsection{A cylindrical vortex in a compressible, infinite and rotating environment:
an Euler-Coriolis case\label{sub:A-cylindrical-vortex-in-a-compressible}}

This illustration describes the evolution of a cylindrical vortex
in an infinite compressible and rotating environment. The amplitudes
and radial dependence of the compressible and rotational fields are
chosen in such a way that if $Y$ is the radius of a large circle
having the vortex at his center, if $M(Y,\lambda,\mu,\omega)$ is
the mass per unit height contained in this cylinder and if $\nu^{2}$
is the mass density in the limit $Y\rightarrow\infty$, then, omitting
exponentially decreasing contributions, we have 
\begin{equation}
M(Y,\lambda,\mu,\omega)/\pi Y^{2}=\nu^{2}+\mathcal{O}(\omega/Y)+\mathcal{O}\left(\left(\lambda Y\right)^{-4}\right),
\end{equation}
where $\mathcal{O}$ denotes the asymptotic notation.

The fields chosen and compatible with the above prescription are
\begin{equation}
\phi_{EC}'(y,\omega)=\frac{\lambda\sqrt{\nu^{2}-\omega^{2}}y^{2}}{\sqrt{1+\lambda^{2}y^{2}}},
\end{equation}
and
\begin{equation}
\psi_{EC}'(y,\omega)=s\frac{\sqrt{\nu^{2}-\omega^{2}}}{\lambda}\left(1-e^{-\mu y}\right),
\end{equation}
the parameter $s=+/-1$ meaning that the amplitude of the rotational
part of the fluid can have both signs.

At this point, let us check the prescription concerning the fields
chosen. We have indeed
\begin{align}
M(Y)\text{/\ensuremath{\pi Y^{2}}}= & ((\phi_{EC}'(Y,\omega))^{2}+\left(\omega Y+\psi_{EC}'(Y,\omega)\right)^{2})Y^{-2}\nonumber \\
= & \left(\left(\frac{\lambda\sqrt{\nu^{2}-\omega^{2}}Y^{2}}{\sqrt{1+\lambda^{2}Y^{2}}}\right)^{2}+\left(\omega Y+s\frac{\sqrt{\nu^{2}-\omega^{2}}}{\lambda}\left(1-e^{-\mu Y}\right)\right)^{2}\right)Y^{-2}\nonumber \\
= & \nu^{2}-\omega^{2}+\omega^{2}+2s\frac{\sqrt{\nu^{2}-\omega^{2}}}{\lambda}\left(1-e^{-\mu Y}\right)\omega Y^{-1}\nonumber \\
 & +\left(\nu^{2}-\omega^{2}\right)\left(\left(1-e^{-\mu Y}\right)^{2}-1/\left(1+\left(\lambda Y\right)^{-2}\right)\right)\left(\lambda Y\right)^{-2},
\end{align}
as claimed. It is worth giving the densities associated to the two
fields. They are
\begin{equation}
\rho_{\phi_{EC}}(y,\omega)=\left(\nu^{2}-\omega^{2}\right)\left(1-\frac{1}{1+\left(\lambda y\right)^{2}}\right),
\end{equation}
and
\begin{equation}
\rho_{\psi_{EC}}(y,\omega)=\frac{\left(\nu^{2}-\omega^{2}\right)\mu}{\lambda^{2}}e^{-\mu y}\frac{\left(1-e^{-\mu y}\right)}{y}.
\end{equation}
\begin{figure}

\begin{centering}
\includegraphics[width=10cm]{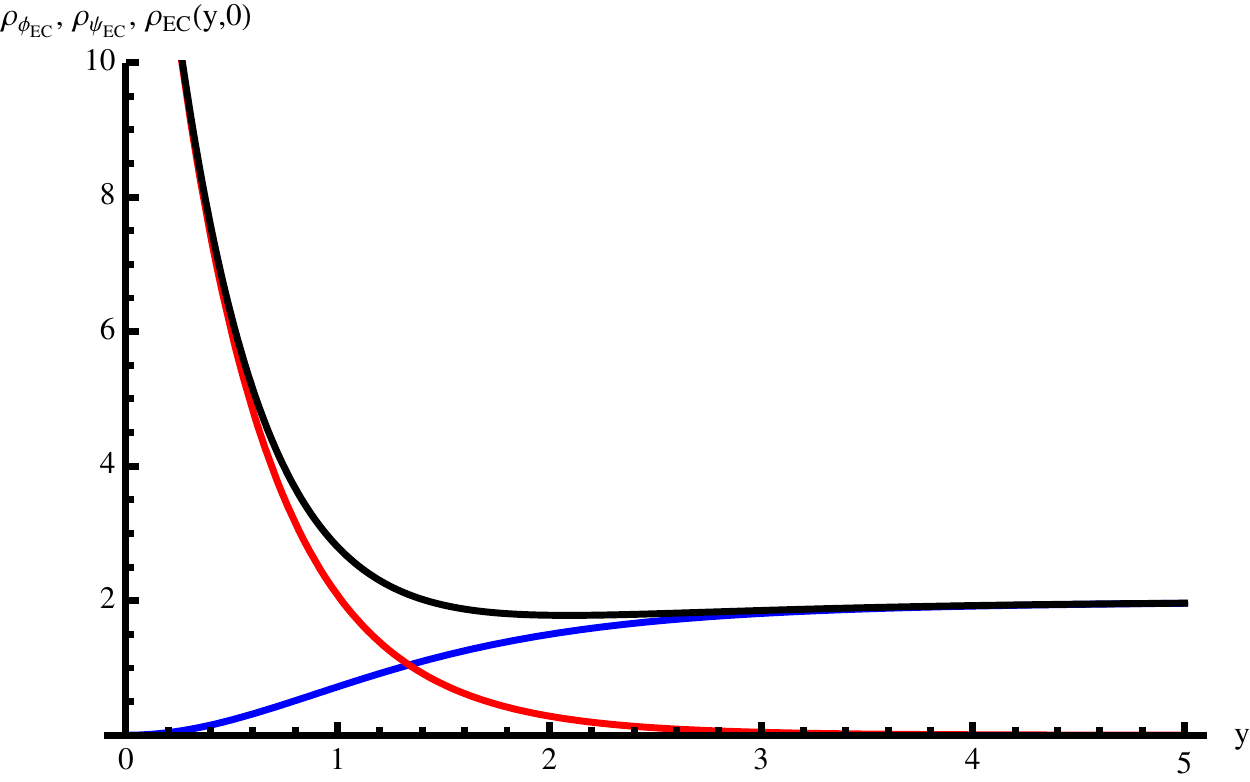}\protect\caption{\label{fig:ecgraph1}Initial densities of the Euler-Coriolis case
at zero frequency.}

\par\end{centering}

\end{figure}
These field densities at zero frequency and their sum are plotted
on Figure \ref{fig:ecgraph1}. We have next
\begin{equation}
\mathbf{v(}y,\omega)^{2}=(\phi_{EC}'(y,\omega))^{2}+\left(\omega y+\psi_{EC}'(y,\omega)\right)^{2},
\end{equation}
and
\begin{equation}
\mathbf{r}(y,\omega,t)^{2}=(y+t\phi_{EC}'(y,\omega))^{2}+t^{2}(\omega y+\psi_{EC}'(y,\omega))^{2}.
\end{equation}
Consider next the regularity conditions. Defining the initial effective
density
\begin{align}
\rho_{EC}(y,\omega)= & \partial\left(\mathbf{v}\mathbf{(y,\omega,s)}^{2}\right)/\partial\left(y^{2}\right)\nonumber \\
= & \omega^{2}+\omega\Delta\psi_{EC}(y,\omega)+\rho_{\phi EC}(y,\omega)+\rho_{\psi EC}(y,\omega),
\end{align}
we have
\begin{equation}
\rho_{EC}(y,\omega)>0,
\end{equation}
and
\begin{equation}
\rho_{EC}(y,\omega,t)=\rho_{EC}(y,\omega)(1+t\Delta\phi_{EC}(y,\omega)+t^{2}\rho_{EC}(y,\omega))^{-1}>0.
\end{equation}
The roots in the time variable of $\rho_{EC}(y,\omega,t)^{-1}=0$
are
\begin{equation}
t_{\pm}(y,\omega)=-\frac{-\Delta\phi_{EC}(y,\omega)\pm D_{EC}(y,\omega)^{1/2}}{2\rho_{EC}(y,\omega)},
\end{equation}
with the discriminant
\begin{equation}
D_{EC}(y,\omega)=\frac{\Delta\phi_{EC}(y,\omega)^{2}-4\rho_{EC}(y,\omega)}{\left(2\rho_{EC}(y,\omega)\right)^{2}}.
\end{equation}
The regularity condition implied is
\begin{equation}
D(y,\omega,\nu,\lambda,\mu.s)<0.
\end{equation}
\begin{figure}

\begin{centering}
\includegraphics[width=10cm]{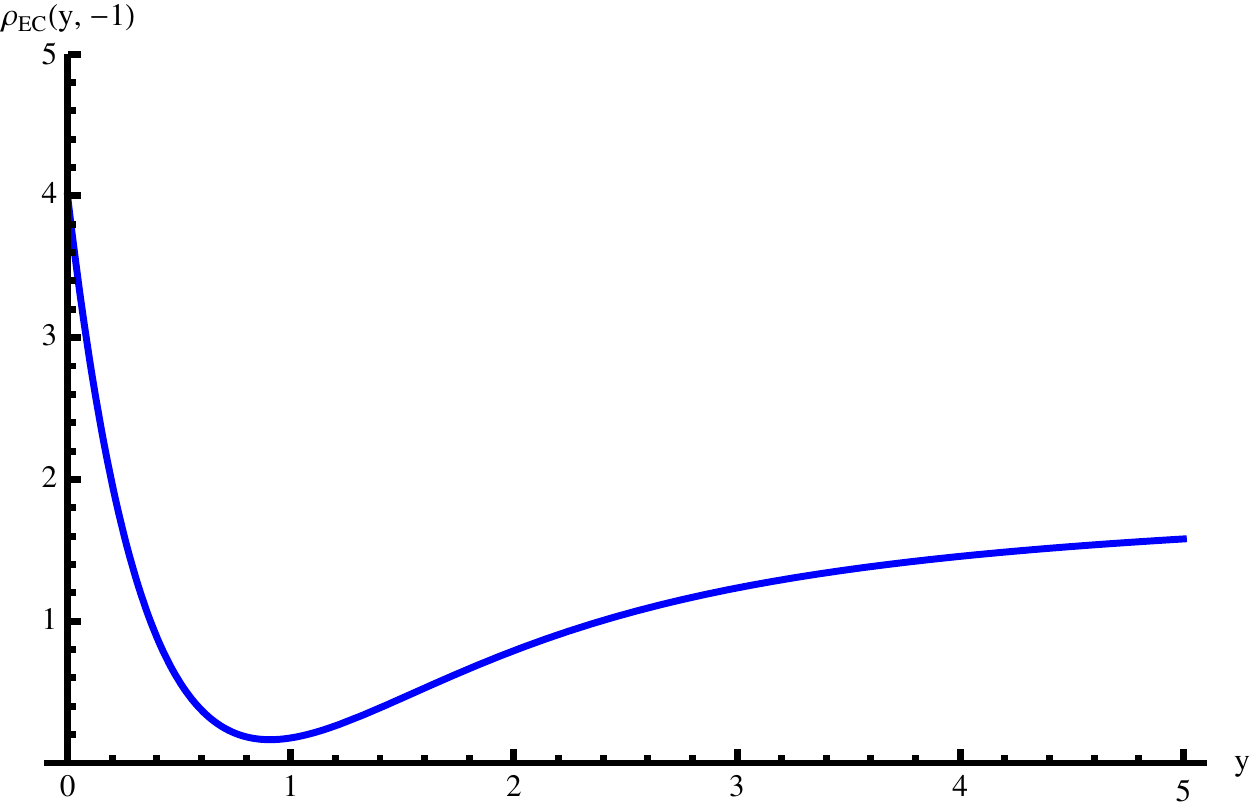}\protect\caption{\label{fig:ecgraph2}Initial density of the Euler-Coriolis case at
$\omega=-1$.}

\par\end{centering}

\end{figure}
This condition ensures that the roots $t_{+,-}(y,\omega,\nu,\lambda,\mu,s)$
be complex conjugate. In summary, there are two conditions: $\rho_{EC}(y,\omega,0,\nu,\lambda,\mu,s)>0$
and $D(y,\omega,\nu,\lambda,\mu.s)<0.$

Focusing on a small 0-dimensional subspace of the complete, 5 dimensional,
parameter space, which exhibits singular solutions, we choose the
following illustration: $\lambda=0.5$, $\mu=1.5$, $\nu^{2}=2$,
$\omega=-1$ and $s=1.$ In Figure \ref{fig:ecgraph2} is plotted
the initial effective density $\rho_{EC}(y,-1,0).$
\begin{figure}

\begin{centering}
\includegraphics[width=10cm]{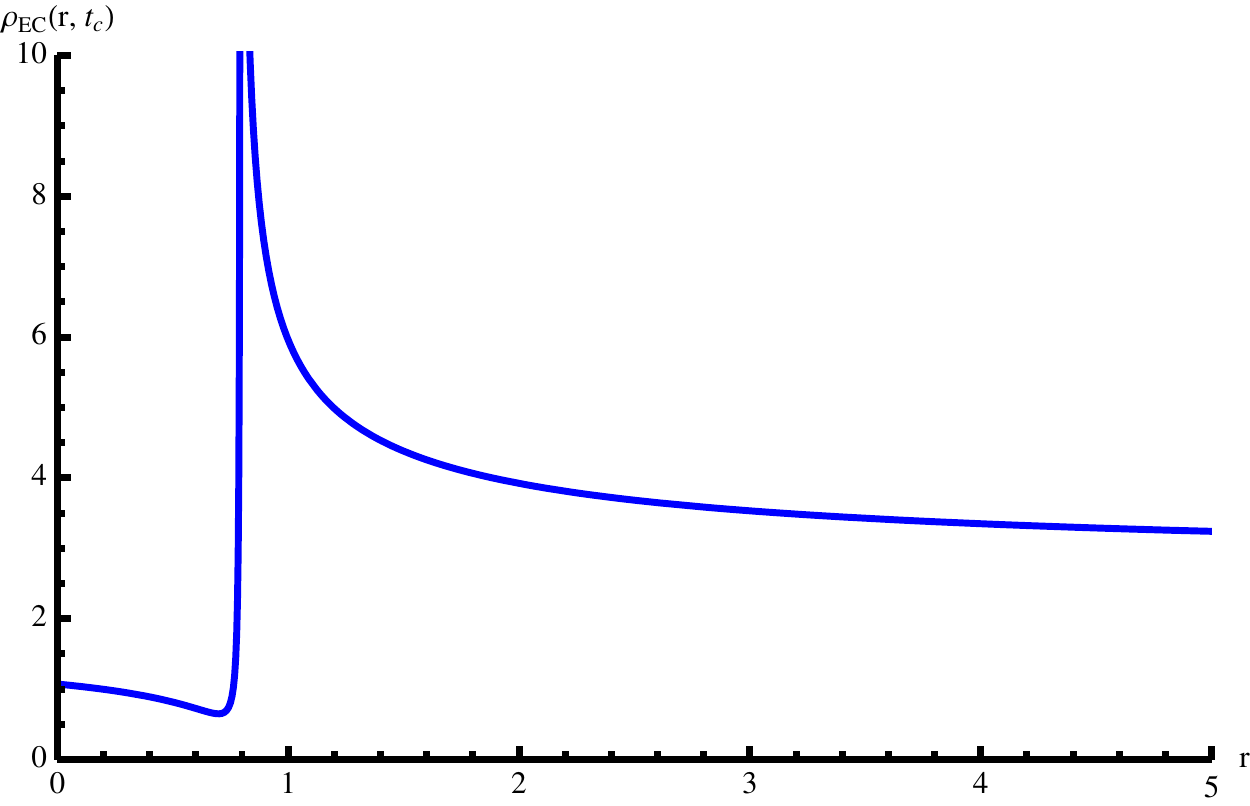}\protect\caption{\label{fig:ecgraph4}Critical density of the Euler-Coriolis case.}

\par\end{centering}

\end{figure}

\begin{figure}
\begin{centering}
\includegraphics[width=10cm]{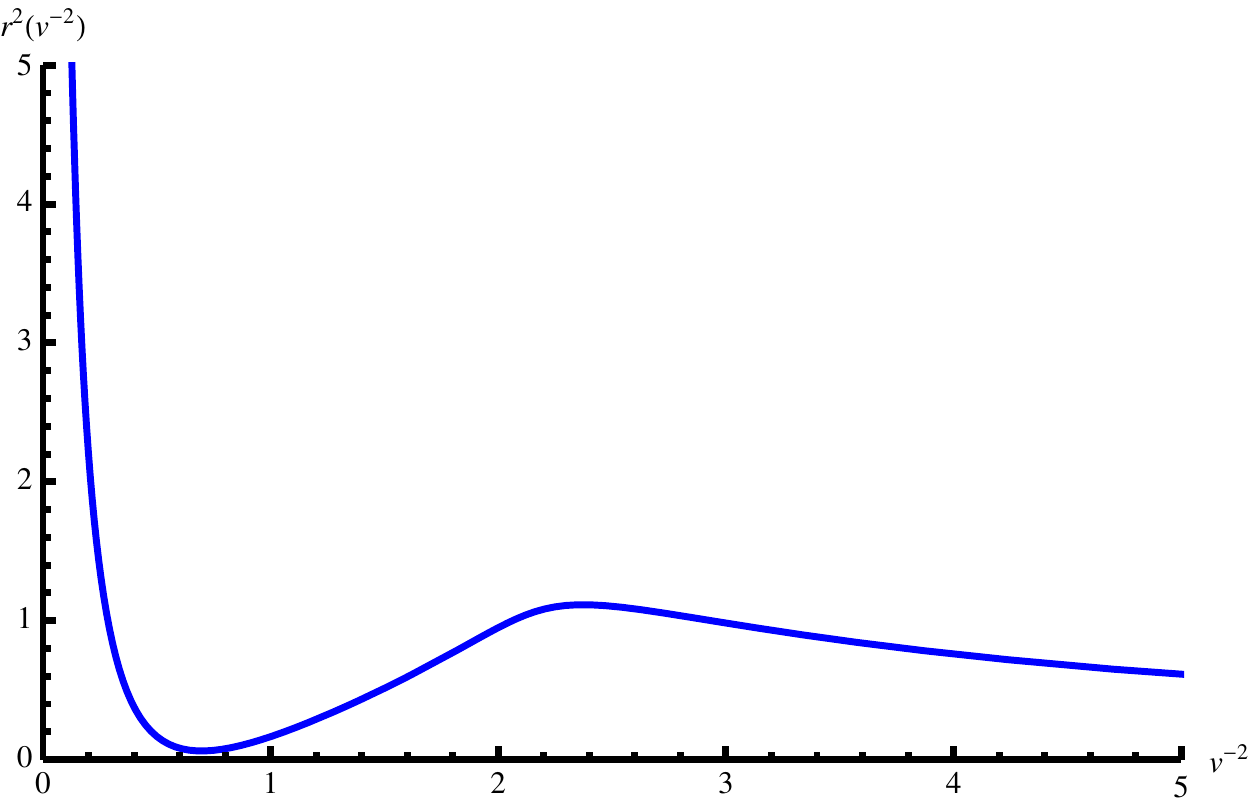}
\par\end{centering}

\protect\caption{\label{fig:ecgraph6}van der Waals analogy of the Euler-Coriolis case.}

\end{figure}

Numerical analysis of the critical density $\rho_{EC}(y,-1,t_{c}),$
shown in Figure \ref{fig:ecgraph4}, of $\rho_{EC}(y,-1,t_{c})^{-1}$,
$\rho_{EC}($ $r,-1.t_{c})$ and $\rho_{EC}(r,-1,t_{c})^{-1}$ gives
$y_{c}\thickapprox1.5$, $r_{c}\thickapprox0.8$ and $t_{c}\thickapprox-0.826.$
For $t>t_{c}$, the solutions are regular. For the sub-critical time
$t_{<}=-1.68<t_{c},$ $\lambda=0.5$, $\mu=1.5$, $\nu^{2}=2$, $s=1$,
$\omega=-1$, for instance, the function $\mathbf{v}^{2}(\mathbf{r}^{2},-1,t_{<})$
and $\mathbf{r}^{2}(\mathbf{v}^{-2},-1,t_{<})$ show a behavior similar
to that of the functions $\mathbf{u}^{2}(\mathbf{x}^{2},t)$ and $\mathbf{x}^{2}(\mathbf{u}^{-2\text{ }},t)$
of the previous illustration. 
\begin{figure}
\begin{centering}
\includegraphics[width=10cm]{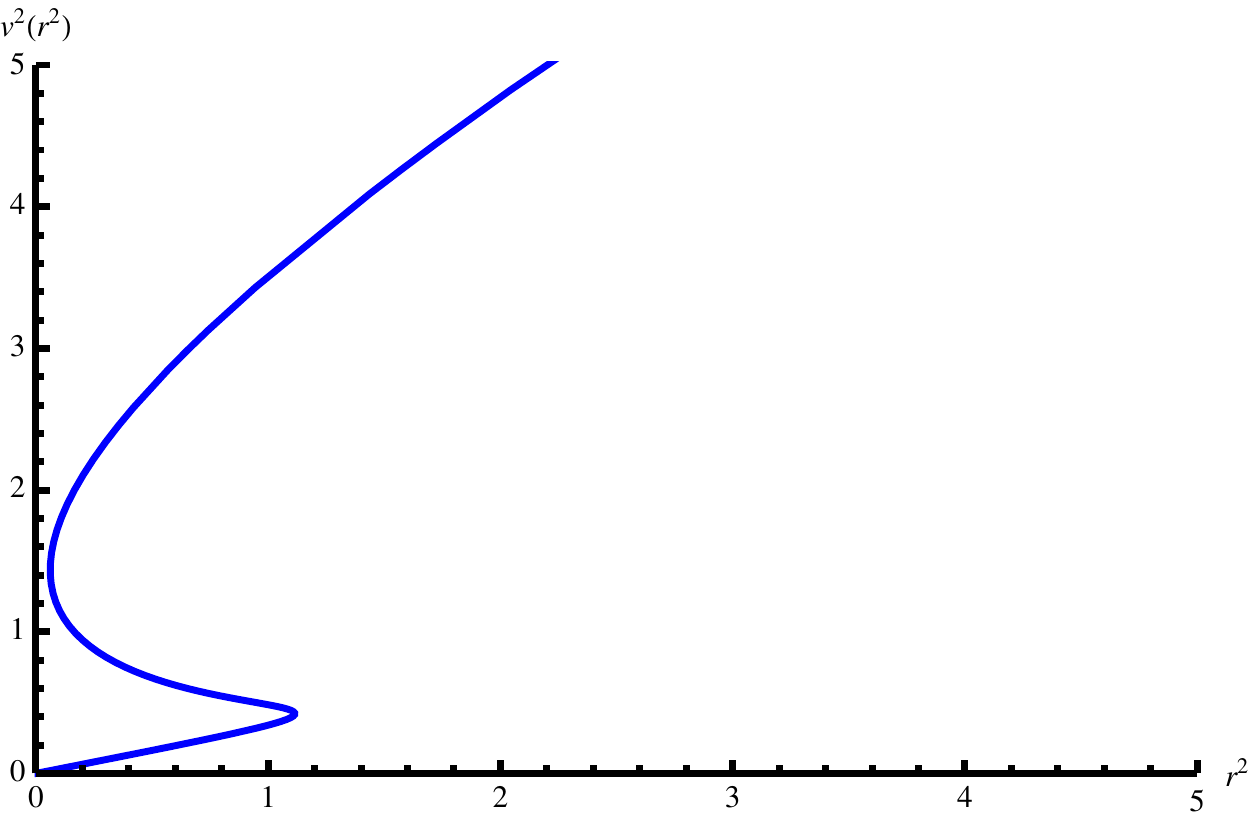}\protect\caption{\label{fig:ecgraph5}Weiss analogy of the Euler-Coriolis case.}

\par\end{centering}

\end{figure}
This is shown in Figure \ref{fig:ecgraph5} and Figure \ref{fig:ecgraph6}. 

\subsection{Variational formulation\label{sub:Variational-formulation}}

The thermodynamical analogies evoked in the subsection \ref{sub:A-cylindrical-vortex-in-a-compressible}
suggest to give the following variational formulation. For the first
illustration, we introduce the potential function:
\begin{equation}
G_{B}(\mathbf{u}^{2},t):=\int_{0}^{\mathbf{u}^{2}}d\mathbf{\widetilde{u}}^{2}\mathbf{x}^{2}(\mathbf{\widetilde{u}}^{2},t),
\end{equation}
and define the Legendre Transform (turns out to be minus a ``Free
Energy'' functional):
\begin{align}
F_{B}(\mathbf{x}^{2},t)= & \sup_{\mathbf{u}^{2}}\left[\mathbf{x}^{2}\mathbf{u}^{2}-G_{B}(\mathbf{u}^{2},t)\right]\nonumber \\
= & \int_{0}^{\mathbf{x}^{2}}d\widetilde{\mathbf{x}}^{2}\mathbf{u}_{\sup}^{2}(\mathbf{\widetilde{x}}^{2},t).\label{eq:F_B variational formula}
\end{align}
\begin{figure}

\begin{centering}
\includegraphics[width=10cm]{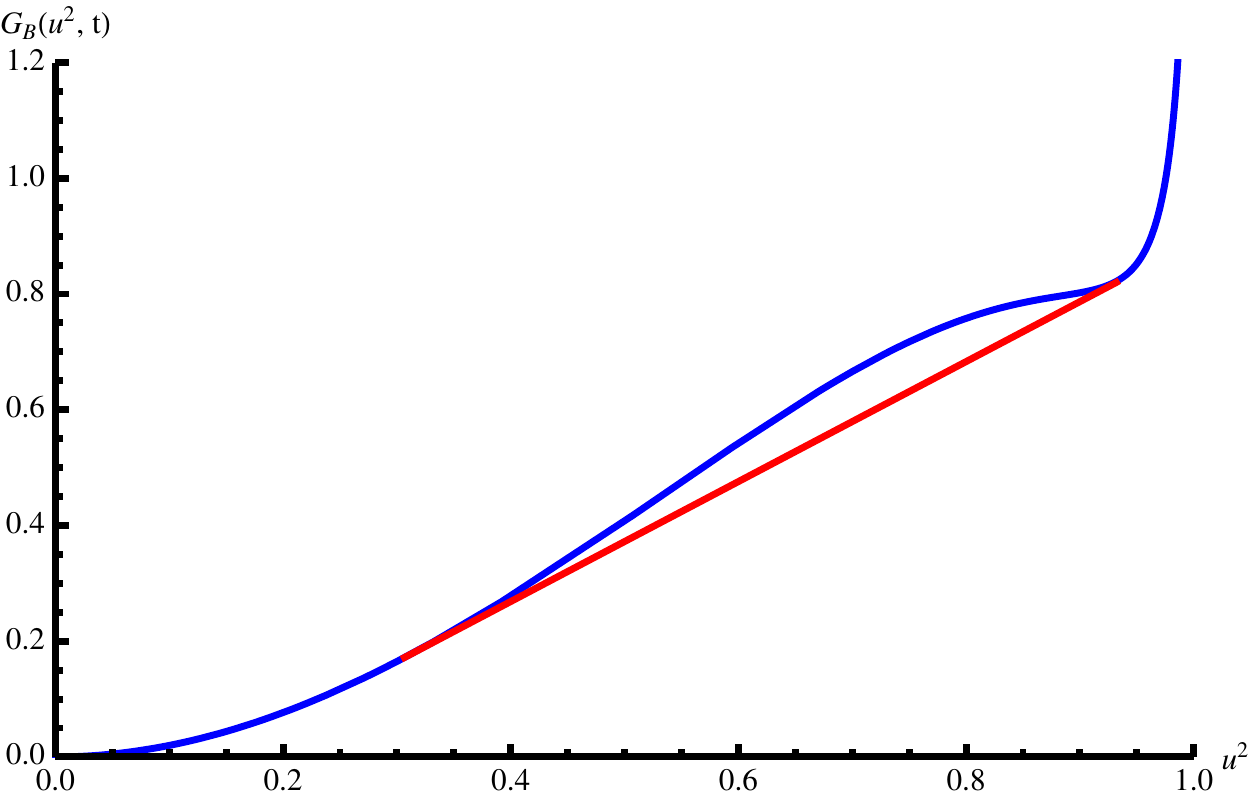}\protect\caption{\label{fig:bgraph8}Convex envelope of the Burgers case.}

\par\end{centering}

\end{figure}

\begin{figure}

\begin{centering}
\includegraphics[width=10cm]{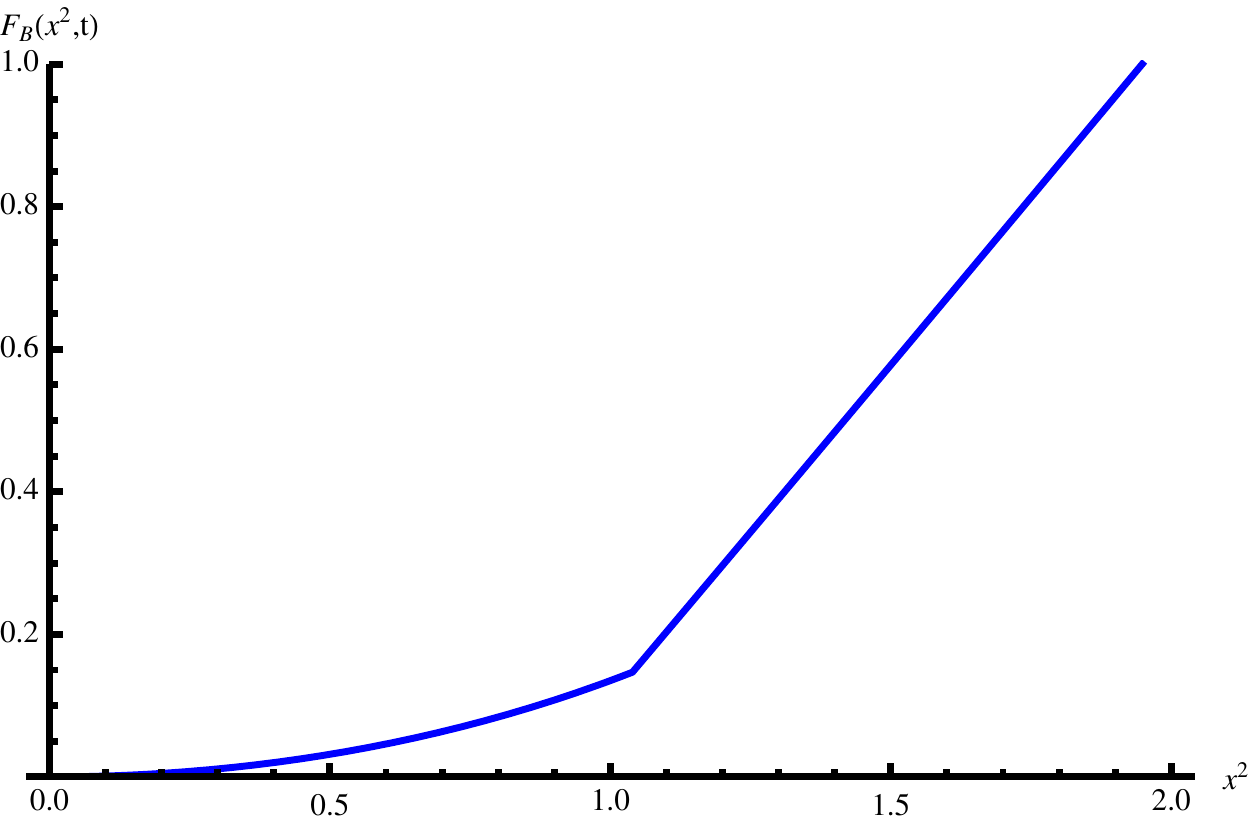}\protect\caption{\label{fig:new2dburgersf}Legendre transform of the Burgers case.}

\par\end{centering}

\end{figure}
\begin{figure}

\begin{centering}
\includegraphics[width=10cm]{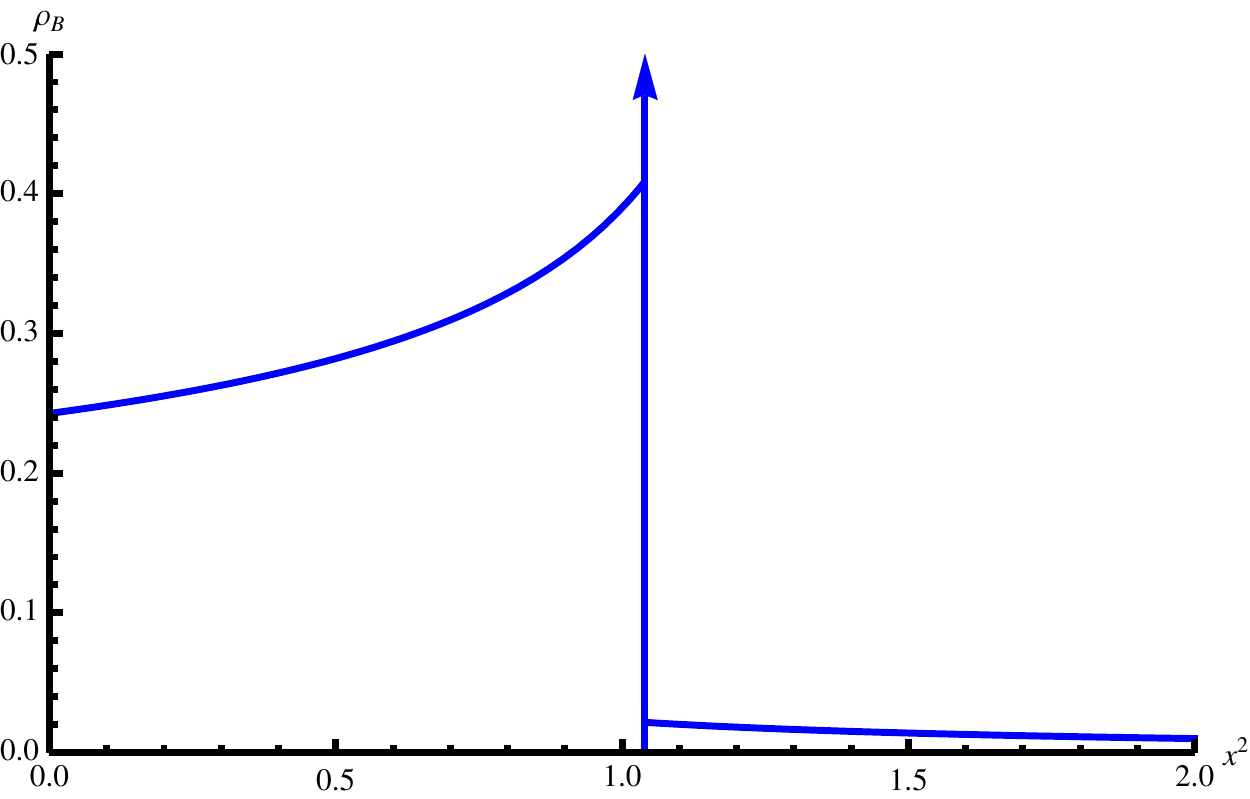}\protect\caption{\label{fig:bgraph10}Sub-critical density of the Burgers case, $t=-3$.}

\par\end{centering}

\end{figure}

This transform implies the construction of the convex envelope of
$G_{B}(\mathbf{u}^{2},t=-3),$ shown in Figure \ref{fig:bgraph8}
and the operation $\sup_{\mathbf{u}^{2}}\left[.\right]$ which generates
the $C_{0}$ function $F(\mathbf{x}^{2},t=-3)$, shown in Figure \ref{fig:new2dburgersf}.
\begin{figure}

\centering{}\includegraphics[width=10cm]{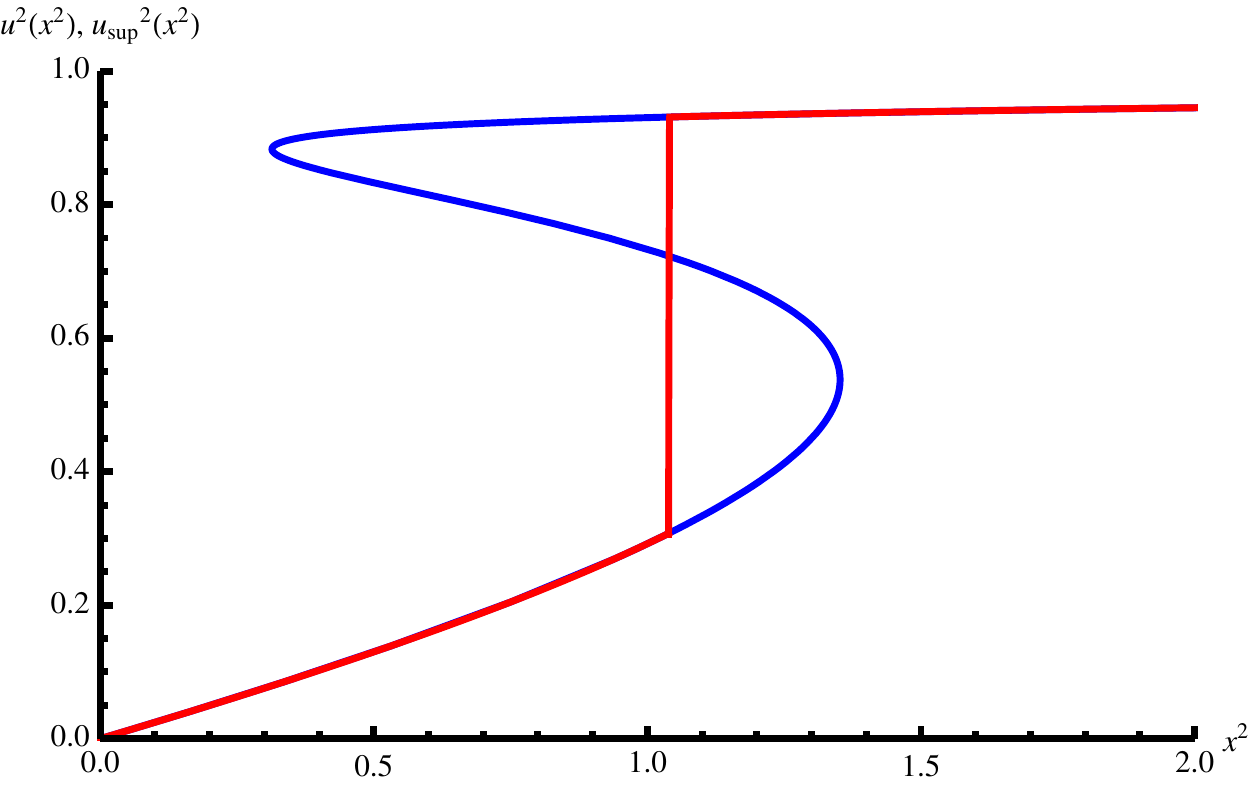}\protect\caption{\label{fig:bgraph9}Maxwell construction of the Burgers case.}
\end{figure}
Then, $\partial F_{B}(\mathbf{x}^{2},t)/\partial\left(\mathbf{x}^{2}\right)=$
$\mathbf{u}_{\sup}^{2}(\mathbf{x}^{2},t)$ introduces a discontinuity,
a vertical cut, in the graph of of $\mathbf{u}^{2}$ versus $\mathbf{x}^{2}$
at $t=-3$, as shown in Figure \ref{fig:bgraph9} and which satisfies
the equal area rule familiar in Thermodynamics. The final result is
a density profile given by $\partial\left(\mathbf{u}_{\sup}^{2}(\mathbf{x}^{2},t)\right)/\partial\left(\mathbf{x}^{2\text{ }}\right)$and
composed of a Dirac distribution and of two tails, as shown in Figure
\ref{fig:bgraph10}.
 If $\mathbf{x}^{\ast2\text{ }}$is the coordinate at separation,
$\mathbf{u}_{\sup,\pm}^{2},$ the corresponding velocities squared
and $\chi(\xi),$ the characteristics function, we have the measure,
sub-critical ($t=-3$), solution
\begin{equation}
\rho_{B}(\mathbf{x}^{2},t)=\rho_{B,-}(\mathbf{x}^{2},t)\chi(\mathbf{x}^{\ast2}-\mathbf{x}^{2})+(\mathbf{u}_{\sup,+}^{\ast2}-\mathbf{u}_{\sup,-}^{\ast2})\delta(\mathbf{x}^{2}-\mathbf{x}^{\ast2})+\rho_{B,+}(\mathbf{x}^{2},t)\chi(\mathbf{x}-\mathbf{x}^{\ast2}).
\end{equation}
It is clear that by construction the total mass is conserved.

For the second illustration we proceed similarly. We define the potential
\begin{equation}
G_{EC}(\mathbf{v}^{2},\omega,t)=\int_{0}^{\mathbf{v}^{2}}d\widetilde{\mathbf{v}}^{2}\mathbf{r}^{2}(\widetilde{\mathbf{v}}^{2},\omega,t),
\end{equation}
and the Legendre transform (another minus of a ``Free Energy'' Functional),
\begin{align}
F_{EC}(\mathbf{r}^{2},\omega,t)= & \sup_{\mathbf{v}^{2}}\left[\mathbf{r}^{2}\mathbf{v}^{2}-G_{EC}(\mathbf{v}^{2},\omega,t)\right]\nonumber \\
= & \int_{0}^{\mathbf{r}^{2}}d\widetilde{r}^{2}\mathbf{v}_{\sup}^{2}(\mathbf{\widetilde{r}}^{2},\omega,t).\label{eq:F_EC variational formula}
\end{align}
\begin{figure}
\begin{centering}
\includegraphics[width=10cm]{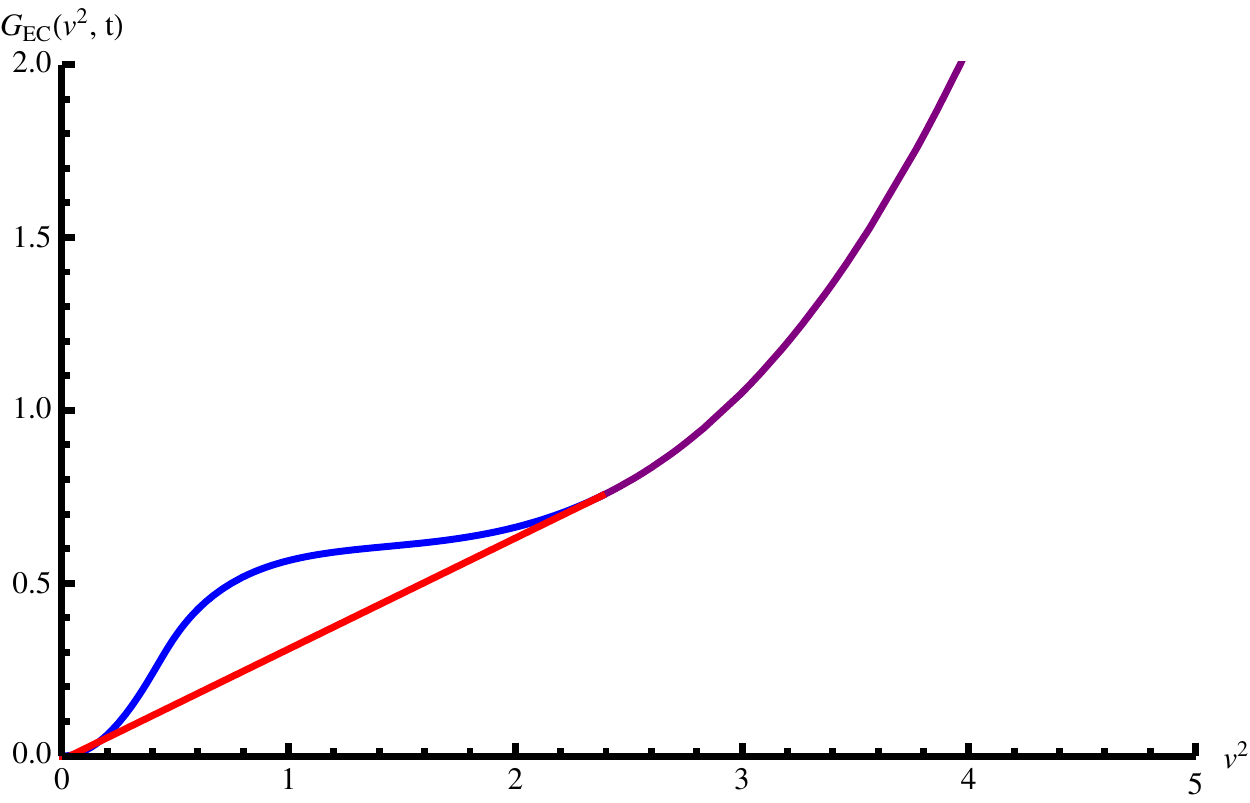}\protect\caption{\label{fig:ecgraph7}Convex envelope of the Euler-Coriolis case.}

\par\end{centering}

\end{figure}
\begin{figure}

\begin{centering}
\includegraphics[width=10cm]{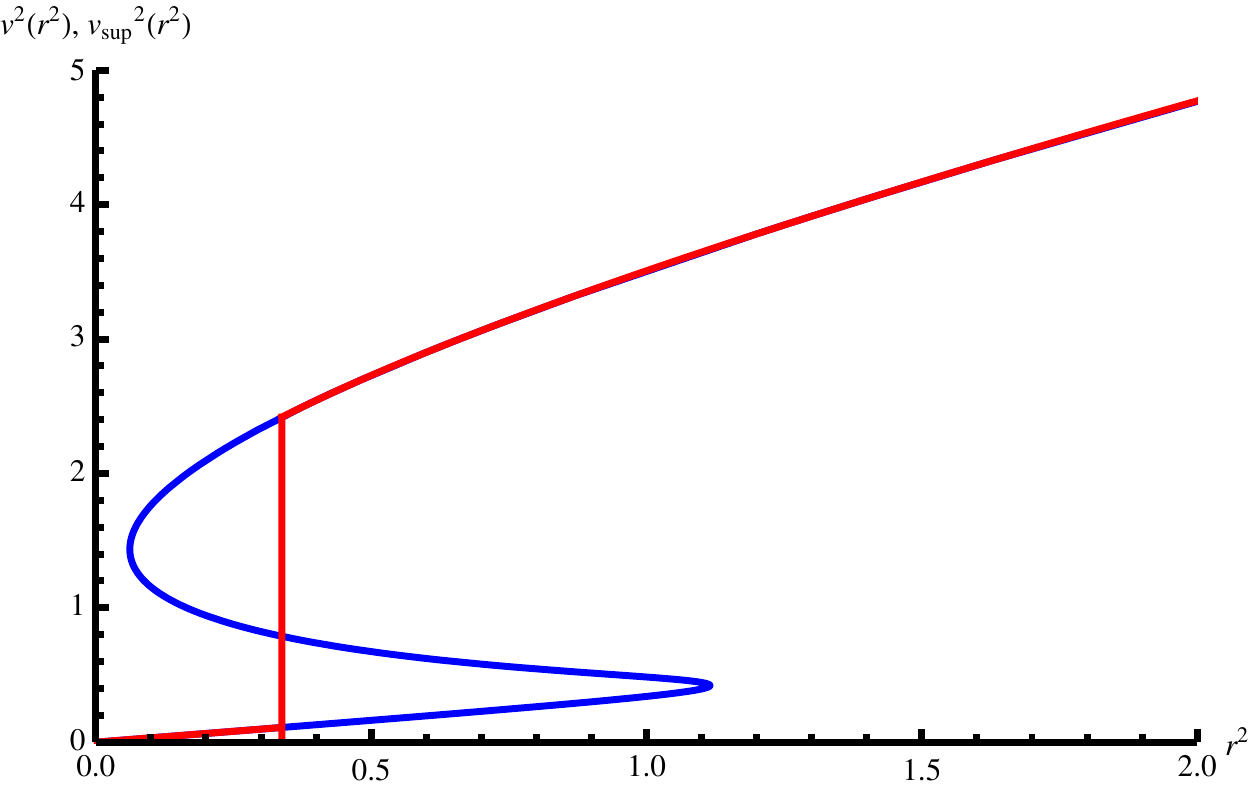}\protect\caption{\label{fig:ecgraph8}Maxwell construction of the Euler-Coriolis case.}

\par\end{centering}

\end{figure}
\begin{figure}

\centering{}\includegraphics[width=10cm]{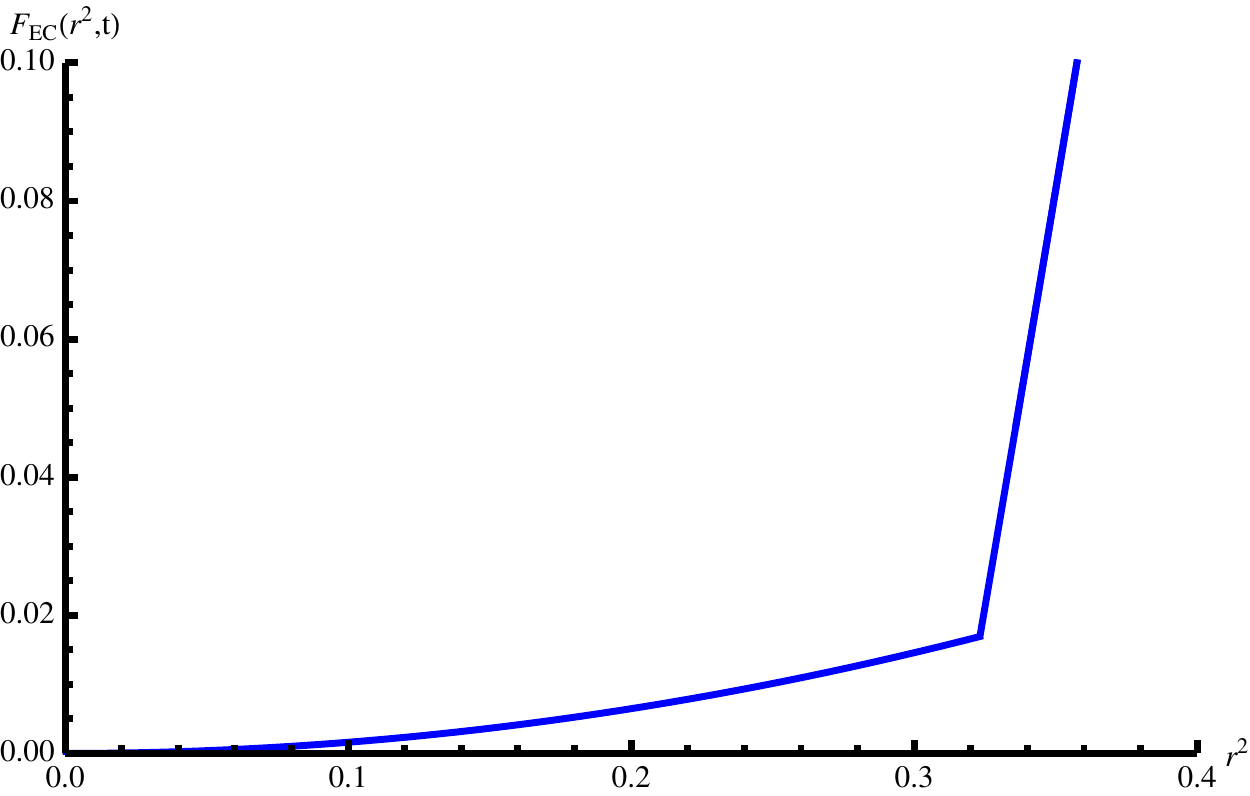}\protect\caption{\label{fig:newstartingfromecgraph8}Legendre transform of the Euler-Coriolis.}
\end{figure}
\begin{figure}

\begin{centering}
\includegraphics[width=10cm]{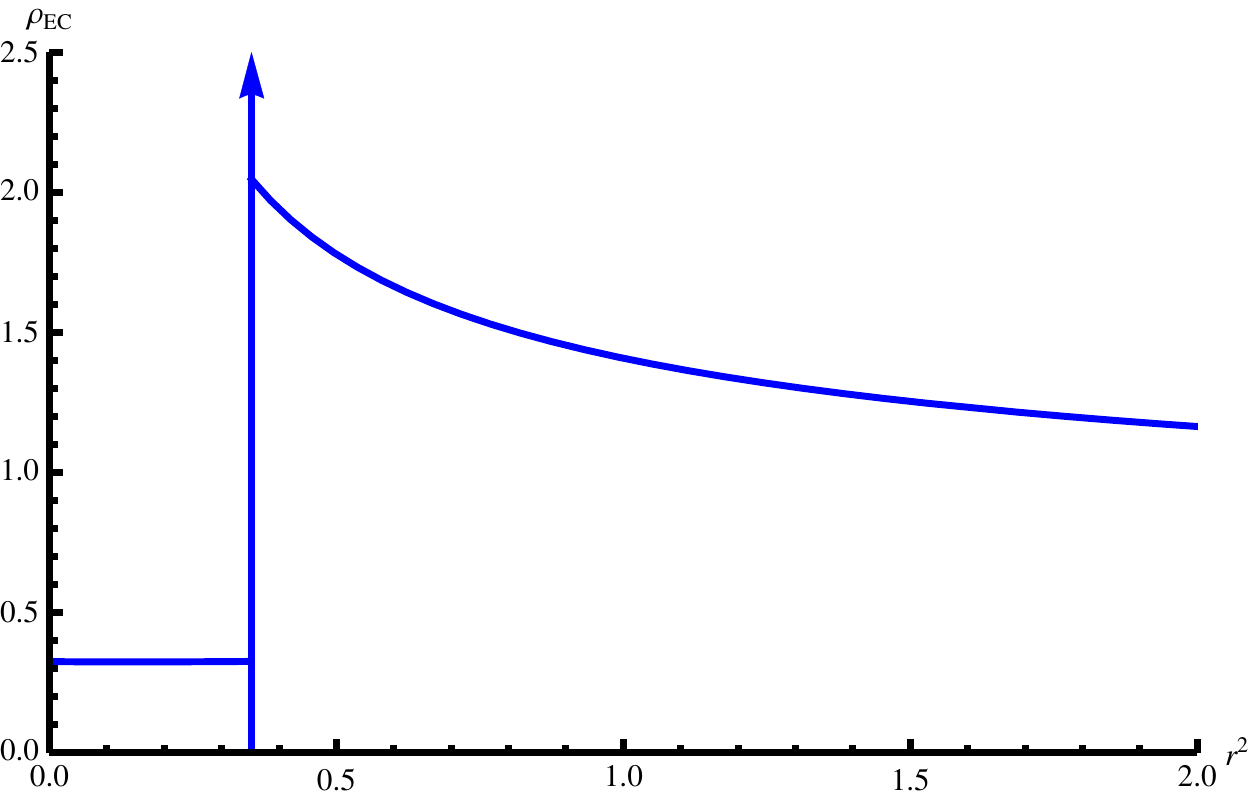}\protect\caption{\label{fig:ecgraph9}Sub-critical density of the Euler-Coriolis case,
$t=-1.68.$}

\par\end{centering}

\end{figure}
We draw $G_{EC}(\mathbf{v}^{2},\omega,t)$ and construct its convex
envelope (Figure \ref{fig:ecgraph7}), the ensuing $\mathbf{v}_{\sup}^{2}(\mathbf{r}^{2},\omega,t)$
 (Figure \ref{fig:ecgraph8}), the $C_{0}$ function $F_{EC}(\mathbf{r}^{2},t,\omega)$,
(Figure \ref{fig:newstartingfromecgraph8}) and the density profile
(Figure \ref{fig:ecgraph9}). With $t_{<}=-1.68$ and $\omega=-1,$
we obtain figures similar but not identical to those of the first
illustration.

\subsection{One-dimensional version of the variational formula}

As a fallout of the content of the Subsection \ref{sub:Variational-formulation}
it is natural to present a one-dimensional version of the isotropic
2D variational formula presented there. For this purpose, consider
the Burgers equation in 1D where $u,x,t\in\mathbb{R}$
\begin{equation}
\partial_{t}u(x,t)+u(x,t)\partial_{x}u(x,t)=0,
\end{equation}
its solution, with initial coordinate $y\in\mathbb{R}$ and velocity
field $u_{0}(y)$
\begin{equation}
x=y+u_{0}(y)t,
\end{equation}
and its implicit solution
\begin{equation}
u(x,t)=u_{0}(x-u(x,t))=u_{0}(y).
\end{equation}
Assume the invertibility condition
\begin{equation}
y=u_{0}^{-1}(u).
\end{equation}
Then
\begin{equation}
x(u,t)=u_{0}^{-1}(u)+tu.
\end{equation}
Our variational formula can now be presented. Indeed, let
\begin{equation}
G_{1}(u,t)=\int_{0}^{u}d\widetilde{u}x(\widetilde{u},t),
\end{equation}
and let 
\begin{equation}
F_{1}(x,t)=\sup_{u}\left(ux-G_{1}(u,t)\right).
\end{equation}
It follows that
\begin{equation}
F_{1}(x,t)=\int_{0}^{x}d\widetilde{x}u_{\sup}(\widetilde{x}\mathbf{,}t),
\end{equation}
is a kind of Maupertuis action per unit mass and the correct, measurable
mass density is
\begin{equation}
\rho_{1}(x,t)=d(u_{\sup}(x,t))/dx.
\end{equation}
An example is given in \citep[sect.4]{ChoMF}, with $u_{0}(y)=\alpha\mathbf{+\tanh}\left(\gamma y\right)$,
$\alpha$ and $\gamma$ being two parameters. This gives 
\begin{equation}
x(u,t)=\frac{1}{2\gamma}\ln\left(\frac{1+u-\alpha}{1-u+\alpha}\right)+tu.
\end{equation}

With a critical time $t_{c}=-1/\gamma$, this example is treated in
details in \citep[p. 850]{ChoMF} where several illustrations are
shown. It is interesting to compare this formulation with that of
Hopf-Lax \citep[Sect. 3, p.847]{ChoMF} and \citep[Sect. 2.3.2.b. p.123]{Eva}.
As expected our variational formulation is equivalent to that of Hopf-Lax
for the one-dimensional problems. As illustration consider the former
example with $\gamma=1$ and $\alpha=0$. Then 
\begin{align}
F_{1}\left(x,t\right) & =\sup_{u}\left(ux-G_{1}(u,t)\right)\nonumber \\
 & =\sup_{u}\left(ux-\frac{tu^{2}}{2}+h_{2}\left(\frac{1+u}{2}\right)-\ln2\right),\label{eq:exemple one-dim F1}
\end{align}
where $h_{2}$ is identified to be the binary entropy function, namely
\begin{equation}
h_{2}\left(p\right):=-p\ln p-\left(1-p\right)\ln\left(1-p\right).
\end{equation}
It is interesting to point out the following analogy with a one dimensional
Ising model treated in the mean-field approximation. Let $J$ be the
coupling constant, $H$ be the magnetic field, $M\left(J,H\right)$
the magnetization and $\Phi\left(J,H\right)$ its free energy. It
is well is known that
\begin{equation}
\Phi\left(J,H\right)=\inf_{m}\left(-Hm+\frac{J}{2}m^{2}-\left(h_{2}\left(\frac{1+m}{2}\right)-\ln2\right)\right),
\end{equation}
and that
\begin{equation}
M\left(J,H\right)=\frac{\partial}{\partial H}\Phi\left(J,H\right).
\end{equation}
It transpires that the analogy reads $J=-t$, $H=x$, $M=-u$ and
$\Phi=-F_{1}$ thus explaining the designation of $F_{1}$ as minus
a ``Free Energy''.

At this point it is worth comparing the variational formula \eqref{eq:exemple one-dim F1}
with the one deriving from the Hopf-Lax principle
\begin{equation}
u_{{\rm HL}}\left(x,t\right)=\frac{\partial}{\partial x}\inf_{y}\left(\frac{\left(x-y\right)^{2}}{2t}+\ln\left(\cosh\left(y\right)\right)\right).
\end{equation}
We observe that
\begin{equation}
\inf_{y}\left(\frac{\left(x-y\right)^{2}}{2t}+\ln\left(\cosh\left(y\right)\right)\right)=\sup_{u}\left(ux-\frac{tu^{2}}{2}+h_{2}\left(\frac{1+u}{2}\right)-\ln2\right),
\end{equation}
as claimed above.

We notice that in higher dimensions the Hopf-Lax formula can be generalized
for compressible systems but not for the combined compressible and
rotational ones, whereas our variational formulas \eqref{eq:F_B variational formula}
and \eqref{eq:F_EC variational formula} does it for radially-symmetric
systems.

\section{Further developments\label{sec:Further-developments}}

If the problems raised in the title of this work is solved in what
concerns the measure solutions of the mass densities of the models
considered, several aspects going beyond its scope have not been touched,
f.i.\ the time evolution of the vorticity and of the dilatation of
these fluids as well as that of their regular and singular spiraling
solutions, the question concerning the applicability of the adhesion
model not speaking of the multi-parametric description of the regimes
displayed by our models.

Further work will imply i) generalization for anisotropic models in
two and three dimensions ii) qualification of entropy solutions with
respect to measure solutions of the 2D compressible and rotational
isotropic models iii) analysis of axis-symmetric flows involving gravitational
forces iv) extension of the strategy of the Jacobian represented densities
for isothermal and isobaric axis-symmetric Burgers and Euler-Coriolis
fluids with Riemann invariants coming into play, and this in view
of meteorological applications.

\section*{Acknowledgment}

The work of M.V. was supported by Swiss National Science Foundation
grant No. 200020-140388.


\bibliographystyle{ieeetr}
\bibliography{Fluid}

\end{document}